\documentclass[letterpaper,twocolumn,10pt]{article}
\usepackage{usenix-2020-09}
\usepackage{hyperref}
\usepackage{xspace}
\usepackage{tikz}
\usepackage{amsmath}
\usepackage{multirow}
\usepackage{subcaption}
\usepackage{algorithm}
\usepackage{algorithmicx}
\usepackage{algpseudocode}

\usepackage{xurl}
\usepackage{xcolor}
\usepackage{ulem}
\algtext*{EndIf}   % Remove "end if" text
\algtext*{EndWhile} % Remove "end while" text
\usepackage{colortbl}
\usepackage{enumitem}
\usepackage{array}
\usepackage{booktabs}
\usepackage{titlesec}
\newcommand\encircle[1]{%
  \tikz[baseline=(X.base)] 
    \node (X) [draw, shape=circle, inner sep=0.5pt] {\footnotesize #1};}

\definecolor{greencolor}{rgb}{0.1,0.7,0.1} % adjust RGB values as needed
\definecolor{orangecolor}{rgb}{1,0.5,0}
\definecolor{redcolor}{rgb}{1,0,0}

\newcommand{\greencircle}{\textcolor{greencolor}{$\bullet$}}
\newcommand{\orangecircle}{\textcolor{orangecolor}{$\bullet$}}
\newcommand{\redcircle}{\textcolor{redcolor}{$\bullet$}}

\newcommand{\conceptname}{MOST\xspace}
\newcommand{\Conceptname}{MOST\xspace}
\newcommand{\conceptnamefull}{Mirror-Optimized Storage Tiering\xspace}

\titlespacing*{\subsubsection}{0pt}{2pt}{2pt}
\titlespacing*{\subsection}{0pt}{2pt}{2pt}
\titlespacing*{\section}{0pt}{2pt}{2pt}
\begin{document}
%don't want date printed

\date{}
% \title{\Large \bf Mirror-Optimized Storage Tiering\\}
\title{\Large \bf Getting the MOST out of your Storage Hierarchy with \\ Mirror-Optimized Storage Tiering}

\author{{\rm Kaiwei Tu, Kan Wu\textsuperscript{\dag}, Andrea C.\ Arpaci-Dusseau, Remzi H.\ Arpaci-Dusseau}\\
University of Wisconsin--Madison, \textsuperscript{\dag}Google 
}

\maketitle
% \vspace{-5em}
%-------------------------------------------------------------------------------

\begin{abstract}
We present \textit{\conceptnamefull} (\conceptname), a novel tiering-based approach optimized for modern storage hierarchies. The key idea of \conceptname is to combine the load-balancing advantages of mirroring with the space-efficiency advantages of tiering. Specifically, \conceptname dynamically mirrors a small amount of hot data across storage tiers to efficiently balance load, avoiding costly migrations. As a result, \conceptname is as space-efficient as classic tiering while achieving better bandwidth utilization under I/O-intensive workloads. We implement \conceptname in Cerberus, a user-level storage management layer based on CacheLib. We show the efficacy of Cerberus through a comprehensive empirical study: across a range of static and dynamic workloads, Cerberus achieves better throughput than competing approaches on modern storage hierarchies especially under I/O-intensive and dynamic workloads.
\end{abstract}

\section{Introduction}

% HIERARCHY: ITS BEEN AROUND, AND STILL IS
Storage hierarchies have been essential to computer system design since
its inception~\cite{Burks+47}. In persistent storage systems, such hierarchies
are commonplace; for example, SSDs are commonly used as a caching layer above
hard drives~\cite{NetappCache}; AutoRAID uses a mirrored set of disks for performance on top of a log-structured RAID-5 layout for capacity~\cite{wilkes1996hp}.

% THE PROBLEM
However, as recent work has shown~\cite{wu2021storage,huang2022ssds}, new device types pose
challenges in the construction of storage hierarchies.  Specifically, new
non-volatile memories~\cite{intel2023optane} and low-latency
SSDs~\cite{intel2023optanessd,kioxia2023ssd} have performance and capacity
profiles that overlap. As such, arranging devices into a strict hierarchy is
difficult~\cite{wu2021storage,huang2022ssds}; doing so ensures that the peak potential of
the storage system is not realized.

% THUS WHAT TO DO?
We are thus left with a significant challenge: how should we manage a
collection of modern devices so as to maximize performance while minimizing
space overheads? As we discuss in this paper, classic techniques fall short.

To understand the issues better, consider a simplified two-device system, with
a faster/smaller ``performance'' device and a slower/larger ``capacity''
device. The first family of approaches we call ``single copy'', as they keep only a single copy of each 
data block in storage. One well-known single-copy approach is 
tiering~\cite{maruf2023tpp,yan2019nimble,Agarwal2017thermostat,dulloor2016heterogeneous,zheng2019ziggurat,zhou2021spitfire,Kwon2017strata,sha2023vtmm,wu2019autoscaling,song2023prism,Raina2023prismDB,vuppalapati2024tiered,raybuck2021hemem, lee2023memtis,yang2017autotiering,guerra2011edt}, which 
(slowly) migrates blocks between tiers based on the hotness or some other optimization goal. Tiering is space-efficient, as
there is one copy of each block; however, a tiering-based system cannot efficiently adapt to changes in load because data need to be migrated around the tiers. Striping~\cite{chen1994raid,salem1986disk} is 
another single-copy approach that statically allocates blocks across devices 
but struggles to manage heterogeneous hierarchies or changes in workload~\cite{cortes2000heterstriping}.

The second family of approaches we call ``multiple copy'', as they replicate some (or all) blocks. 
A well-known multiple-copy approach is caching~\cite{oneil1993lru,wu2021storage,intel2023opencas,forney2002storage,oh2012caching,berg2020cachelib,yang2020analysis,qiu2023frozen,mcallister2021kangaroo}, which
makes copies of popular blocks in the performance device. As a result,
caching wastes capacity (as all copy-based approaches
do); in addition, caching does not utilize the bandwidth of the capacity layer,
thus falling short of peak performance. Mirroring~\cite{Bitton+88-Shadow,chen1994raid}, 
a technique that replicates data across two devices, effectively balances 
the load for read requests by directing them to either device, yet also incurs 
a significant capacity cost.
   
The state-of-the-art approaches for handling heterogeneous memory and storage devices focus on tiering~\cite{vuppalapati2024tiered, chou2017batman}; at a high level, these previous works attempt to maximize the total throughput delivered by all devices by dynamically placing and migrating data across devices such that accesses to each device are proportional to their bandwidth or equalize access latency. In this paper, we introduce \textit{\conceptnamefull} (MOST) and show that mirroring a small amount of hot data across devices, in combination with tiering, greatly improves performance and reduces device writes.  Given these mirrored data, the host can dynamically route requests to different devices to effectively balance load instead of performing costly data migrations across devices like other approaches.

We find that adding a small amount of mirroring is advantageous to pure tiering for three primary reasons. First, mirrors enable the system to quickly react to workload changes by adjusting routing; pure tiering, on the other hand, must slowly migrate a significant amount of data to adjust load distribution. Second, mirrors reduce overall device writes when handling dynamic workloads; migration-based approaches require extensive device writes to move data across tiers. Third, mirroring is more robust to fluctuations in device performance and prevents overreacting with unnecessary migrations. This effect is particularly pronounced for storage devices that suffer from poor tail latency or write-heavy workloads that frequently trigger background activity inside the SSD.

\Conceptname is a novel approach optimized for the modern storage hierarchy. \Conceptname mirrors a small amount of hot data within the classic tiering system, enabling flexible load distribution control. Under low load, \conceptname functions similarly to a classic tiering system, routing requests for mirrored data to the performance device to maximize its utilization. Under high load, \conceptname dynamically routes a portion of requests to the capacity copy of the mirrored data, thereby utilizing the bandwidth of the capacity device more effectively. Key to the success of \conceptname are the algorithms that manage its behavior. \conceptname\ is based on a simple, robust, and self-adjusting optimization mechanism that requires no prior workload knowledge and is capable of handling the varied performance characteristics of different storage hierarchies.

We implement \conceptname as a storage management layer named Cerberus within the widely-used flash cache library, CacheLib~\cite{berg2020cachelib, mcallister2021kangaroo, meta2023cachelib}. Our experiments are conducted on two storage hierarchies using three different real device types~\cite{intel2023optanessd, samsung2023nvme, samsung2023sata, dellnvme, nvmeof}. For static cache workloads, \conceptname achieves up to 2.34$\times$ higher throughput and reduces P99 latency by up to 75\% compared to state-of-the-art systems (Colloid~\cite{vuppalapati2024tiered}, HeMem~\cite{raybuck2021hemem}, and Orthus~\cite{wu2021storage}). Across four production cache workloads, \conceptname increases throughput by up to 1.86$\times$ and reduces P99 GET latency by up to 90\% compared to these systems. Overall, Cerberus enhances throughput under intensive workloads and reduces writes by up to 84\% compared to Colloid under dynamic workloads through its judicious use of mirroring across storage tiers.

\section{Motivation and Background}

In this section, we discuss some of the trends in modern memory and storage hierarchies containing multiple heterogeneous devices. We then describe existing management techniques -- such as striping, tiering, mirroring, and caching-- and analyze their limitations.  We conclude by discussing some of the extra challenges of managing storage hierarchies, as opposed to memory hierarchies.

\subsection{Multi-Device Hierarchies}
\label{hierarchy-intro}
Multiple devices~\cite{zheng2019ziggurat,ren2025polystore,raybuck2021hemem,lee2023memtis,vuppalapati2024tiered,wu2021storage,chou2017batman} are commonly used to serve memory and storage load. Since new devices may be added incrementally over time to increase capacity or performance, these new devices are likely to exhibit different performance characteristics compared to older devices in the system. One natural approach has been to organize memory or storage devices into a {\it hierarchy} of tiers, with newer, faster devices placed as a cache above slower, possibly larger devices; such hierarchies have been essential to systems since their inception~\cite{Burks+47}.

% THE PROBLEM
However, as recent work has shown~\cite{wu2021storage,huang2022ssds}, new device types pose
challenges to ordered hierarchies.  To simplify our discussion, we focus on a two-layered hierarchy containing two types of devices: a {\it performance} device and a {\it capacity} device, where the performance device is more expensive, smaller in capacity, and faster, whereas the capacity device is cheaper, larger, and slower.   In a traditional hierarchy (e.g., DRAM on HDD), the performance gap between the two devices is substantial and the performance of the capacity device can be ignored.  However, with emerging technologies such as non-volatile memory~\cite{intel2023optane}, low-latency SSDs~\cite{intel2023optanessd}, NVMe Flash SSDs~\cite{samsung2023nvme, dellnvme}, SATA Flash SSDs~\cite{samsung2023sata}, and remote storage with NVMeoF~\cite{nvmeof}, disaggregated SSDs~\cite{klimovic2016flash, klimovic2017reflex}, and EBS~\cite{amazonebs, zhangebs}, the performance and capacity devices have overlapping characteristics.  As such, arranging these devices into a strict hierarchy~\cite{wu2021storage,huang2022ssds} limits their potential.

\if 0
Specifically, current 
and low-latency
SSDs~\cite{intel2023optanessd,kioxia2023ssd} have performance and capacity
profiles that overlap. 
\fi

% % THUS WHAT TO DO?
% We are thus left with a significant challenge: how should we manage a
% collection of modern devices so as to maximize performance and capacity utilization? 

\begin{table}[t]

\resizebox{\columnwidth}{!}{%
\begin{tabular}{l|cc|cc|cc}
\multicolumn{1}{c|}{\multirow{2}{*}{\bf \Large Storage Device}} & \multicolumn{2}{c|}{Latency ($\mu$s)} & \multicolumn{2}{c|}{Read (GB/s)} & \multicolumn{2}{c}{Write (GB/s)} \\ %\cline{3-8} 
  & \multicolumn{1}{c|}{4K}    & 16K  & \multicolumn{1}{c|}{4K}    & 16K & \multicolumn{1}{c|}{4K}         & 16K      \\ \hline
Optane SSD                & \multicolumn{1}{c|}{11}    & 18   & \multicolumn{1}{c|}{2.2}   & 2.4 & \multicolumn{1}{c|}{2.2}        & 2.2      \\
PCIe 4.0 NVMe Flash SSD  & \multicolumn{1}{c|}{66}    & 86   & \multicolumn{1}{c|}{1.5}   & 3.3 & \multicolumn{1}{c|}{1.9}        & 2.3      \\
PCIe 3.0 NVMe Flash SSD  & \multicolumn{1}{c|}{82}    & 90   & \multicolumn{1}{c|}{1.0}   & 1.6 & \multicolumn{1}{c|}{1.5}        & 1.6      \\
PCIe 4.0 NVMe Flash SSD over RDMA   & \multicolumn{1}{c|}{88}    & 114  & \multicolumn{1}{c|}{1.2}   & 2.7 & \multicolumn{1}{c|}{1.7}        & 2.3  \\
SATA Flash SSD            & \multicolumn{1}{c|}{104}   & 146  & \multicolumn{1}{c|}{0.38}  & 0.5 & \multicolumn{1}{c|}{0.38}       & 0.5      
\end{tabular}
}
\caption{\textbf{Device Performance.} \textit{Latency measured for a single-thread read workload; bandwidth for a 32-thread workload. Remote device is connected via a 25 Gbps link.}}

\label{tab:device_stat}
\vspace{-0.1in} 
\end{table}

Table \ref{tab:device_stat} shows that latency and bandwidth can be similar for the performance and capacity devices given current technology. Memory and storage systems can be composed of any pair of these devices. Importantly, the performance ratio across the two tiers may be close: the bandwidth ratio for 16KB reads is only 1.5:1 between Optane and PCIe 3.0 NVMe devices and only 1.25:1 between local and remote PCIe 4.0 NVMe devices.  More subtly, the performance ratios are not static and strongly depend on workload; for example, given 4KB reads, the ratio between Optane and PCIe 3.0 NVMe devices increases to approximately 2.2:1.  These performance ratios depend on access sizes, percentage of writes, and concurrency~\cite{wu2019contract}.

\begin{table*}[t!]
\centering
\resizebox{\textwidth}{!}{%
\begin{tabular}{cc|c|c|c|c|c|c|c}

\multicolumn{2}{c|}{} & \multicolumn{4}{c|}{Single Copy} & \multicolumn{3}{c}{Multiple Copy} \\ \hline
\multicolumn{2}{c|}{} & Striping & HeMem & BATMAN & Colloid & Mirroring & Orthus & MOST \\ \hline

\multirow{4}{*}{\textbf{Bandwidth Utilization}} 
& Random Read-only      & Low {\redcircle}  & Low {\redcircle} & Medium {\orangecircle}{\orangecircle}  
                 & Medium {\orangecircle}{\orangecircle} & High {\greencircle}{\greencircle}{\greencircle} 
                 & High {\greencircle}{\greencircle}{\greencircle} & High {\greencircle}{\greencircle}{\greencircle}  \\ \cline{2-9} 

& Random Write-only     & Low {\redcircle}  & Low {\redcircle} & Medium {\orangecircle}{\orangecircle}  
                 & Low {\redcircle}  & Low {\redcircle}  
                 & Low {\redcircle}  & High {\greencircle}{\greencircle}{\greencircle}  \\ \cline{2-9} 

& Random RW-mixed        & Low {\redcircle} & Low {\redcircle} & Medium {\orangecircle}{\orangecircle} 
                 & Medium {\orangecircle}{\orangecircle} & Medium {\orangecircle}{\orangecircle}   
                 & Medium {\orangecircle}{\orangecircle} & High {\greencircle}{\greencircle}{\greencircle}  \\ \cline{2-9}

& Sequential Write     & Low {\redcircle}  & Low {\redcircle} & Low {\redcircle} 
                 & Low {\redcircle} & Low {\redcircle} 
                 & Low {\redcircle}  & High {\greencircle}{\greencircle}{\greencircle}  \\ \hline

\multicolumn{2}{c|}{\textbf{Capacity Utilization}}  
                 & High {\greencircle}{\greencircle}{\greencircle}  
                 & High {\greencircle}{\greencircle}{\greencircle}   
                 & High {\greencircle}{\greencircle}{\greencircle}  
                 & High {\greencircle}{\greencircle}{\greencircle}  
                 & Low {\redcircle}  
                 & Low {\redcircle}  
                 & High {\greencircle}{\greencircle}{\greencircle}  \\ \hline

\multicolumn{2}{c|}{\textbf{Dynamic Workload}}  
                 & Low {\redcircle}  
                 & Low {\redcircle}  
                 & Low {\redcircle} 
                 & Low {\redcircle} 
                 & Medium {\orangecircle}{\orangecircle}   
                 & Medium {\orangecircle}{\orangecircle}  
                 & High {\greencircle}{\greencircle}{\greencircle}  \\ 

\end{tabular}
}
\caption{\textbf{Qualitative Comparison of Different Techniques in a Modern Storage Hierarchy.} 
\textit{Bandwidth utilization is categorized as follows:  Low means no load-balancing mechanism; Medium represents limited load-balancing with restrictions under specific workloads; High indicates effective load-balancing across a wide range of workloads.}}
\label{tab:comparison}
\vspace{-0.1in}
\end{table*}

\subsection{Existing Approaches}
\label{sec:existing_approach}

At a high level, classic approaches for managing multi-device memory and storage can be grouped into two categories: approaches that maintain a single copy of each block (i.e., striping, tiering, and exclusive caching), and approaches that may maintain multiple copies of each block (i.e., mirroring, caching).  We describe these approaches and highlight their inefficiencies, as summarized in Table~\ref{tab:comparison}.

\noindent\textbf{Single Copy.}  {\it Striping}~\cite{patterson1988case, chen1994raid, cortes2000heterstriping, salem1986disk, gafsi1999data, ozden1996disk} allocates a single copy of each block in a predetermined static pattern across devices.  Striping does not handle heterogeneous  devices well: if data is striped evenly across devices, throughput is bottlenecked by the slowest device; if data is striped in a weighted pattern with more data on faster devices, the appropriate weight depends on the workload; furthermore, the ideal striping ratio for performance is unlikely to match the ratio of the devices' capacity (i.e., faster devices are usually smaller).  Thus, striping is not a good match for heterogeneous storage.

\newcommand{\shortref}[1]{\S\ref{#1}}

{\it Tiering}~\cite{maruf2023tpp, yan2019nimble, Agarwal2017thermostat, dulloor2016heterogeneous, zheng2019ziggurat, zhou2021spitfire, Kwon2017strata, sha2023vtmm, wu2019autoscaling, song2023prism, Raina2023prismDB} also allocates a single copy of each block; however, instead of allocating blocks in a static pattern, tiering dynamically places data blocks based on their hotness~\cite{yan2019nimble, Agarwal2017thermostat, zheng2019ziggurat, zhou2021spitfire, Kwon2017strata, song2023prism, Raina2023prismDB, maruf2022multi} or some other optimization goal~\cite{guerra2011edt, yang2017autotiering, chou2017batman}.  Table~\ref{tab:comparison} summarizes the characteristics of three tiering approaches. 
 
First, classic hotness-based tiering is exemplified by HeMem~\cite{raybuck2021hemem}, which manages a memory system containing DRAM and NVM.  HeMem promotes hot data into the performance tier and serves hot data exclusively from the performance tier, leading to inefficient utilization of the capacity tier’s bandwidth; thus, HeMem delivers low bandwidth for intensive workloads due to not effectively utilizing NVM (as shown later \shortref{sec:eval}). Second, BATMAN~\cite{chou2017batman} balances tier bandwidth by migrating data, but its fixed bandwidth ratio prevents it from adapting to hierarchies whose performance ratios vary with workload or evolve over time. Third, Colloid~\cite{vuppalapati2024tiered} also dynamically migrates data between memory tiers to utilize the bandwidth of the capacity device; however, Colloid does not handle dynamic and time-varying workloads because costly data migrations are required to adjust the load distribution when the workload changes (\shortref{sec:micro_dynamic}). 

Other tiering systems, such as AutoTiering~\cite{yang2017autotiering} and EDT-DTM~\cite{guerra2011edt}, also balance data placement to utilize the bandwidth of all tiers, but since they rely solely on data migration to adjust load distribution, their adaptability and responsiveness to workload changes is limited. {\it Exclusive caching}~\cite{gill2008multi} maintains only a single copy of data within the hierarchy: when hot data is promoted or demoted between tiers, the original copy is discarded.  At a high level, exclusive caching is similar to hotness-based tiering but moves data at smaller time intervals; consequently, it behaves similarly. In summary, as shown in Table~\ref{tab:comparison}, single-copy approaches may partially utilize the bandwidth of the capacity device by migrating data across tiers. However, this migration-based strategy struggles to adapt to dynamic workloads, as migration is both costly and slow. In addition to limited responsiveness, migration-based approaches also suffer from excessive device write and performance degradation due to migration-induced traffic (see~\shortref{sec:storage_vs_mem} for further discussion).

\noindent\textbf{Multiple Copies.}  To utilize both devices, other approaches maintain two copies of at least some data blocks across both the performance and capacity devices.  {\it Mirroring}~\cite{Bitton+88-Shadow, chen1994raid} simply replicates all data on both devices.  Mirroring delivers high read bandwidth since reads can be dynamically load balanced across both devices to account for performance differences; however, mirroring delivers low write bandwidth since both copies must be updated and bandwidth is limited by the slower device.  Mirroring also has low capacity utilization since each block is stored on both devices.

{\it Inclusive caching} allocates all items on the capacity device, but replicates only frequently-accessed items on the performance device~\cite{oneil1993lru, wu2021storage, intel2023opencas, forney2002storage, oh2012caching, berg2020cachelib, yang2020analysis, qiu2023frozen, mcallister2021kangaroo}; inclusive caching inherently wastes the capacity of the performance device and fails to exploit the performance of the capacity device~\cite{kim2018sib,ahmadian2019lbica}.  To utilize the performance of the capacity device, Orthus~\cite{wu2021storage} introduces Non-Hierarchical Caching (NHC).  Orthus dynamically redirects read traffic from the performance device to the capacity device when the performance device is overloaded. However, Orthus is built on a traditional caching model, which introduces two fundamental limitations. First, it is space-inefficient—NHC uses the entire capacity of the performance device to store duplicate copies of data from the capacity tier, wasting the performance tier’s storage capacity. Second, Orthus struggles with write-intensive workloads, requiring clean copies for routing read requests: writes that hit in the cache can be handled with write-through or write-back/write-around.  With write-through, both copies remain clean, but additional writes are incurred, and performance is constrained by the write bandwidth of the capacity device.  With write-back/write-around to only one of the block's copies, Orthus can only route subsequent reads to the dirty block. Thus, Orthus performs poorly with writes (\shortref{sec:micro_static}).

Nomad~\cite{xiang2024nomad} proposes a variant to hotness-based tiering that maintains temporary copies of data during migration; specifically, the original copy on the capacity device can still be accessed while the data is being migrated to the performance device.  While Nomad improves the performance penalty of migration, it does not maximize the bandwidth of the underlying devices in the common case.

\textbf{Summary.} Existing approaches for handling heterogeneous memory and storage fail along important metrics (Table~\ref{tab:comparison}). Approaches that maintain a single copy of data, such as striping, tiering (e.g., HeMem, BATMAN, Colloid, AutoTiering, and EDT-DTM), and exclusive caching, fall short in handling dynamic workloads. Current approaches, such as mirroring, inclusive caching, and non-hierarchical caching, that maintain multiple copies of data items suffer from low capacity utilization and struggle with write-intensive workloads. We will show that MOST delivers high bandwidth and reduces device writes across a wide range of static and dynamic workloads with only a small amount of mirrored data.
\begin{figure}[t!]
    \centering
    \includegraphics[width=0.97\columnwidth]{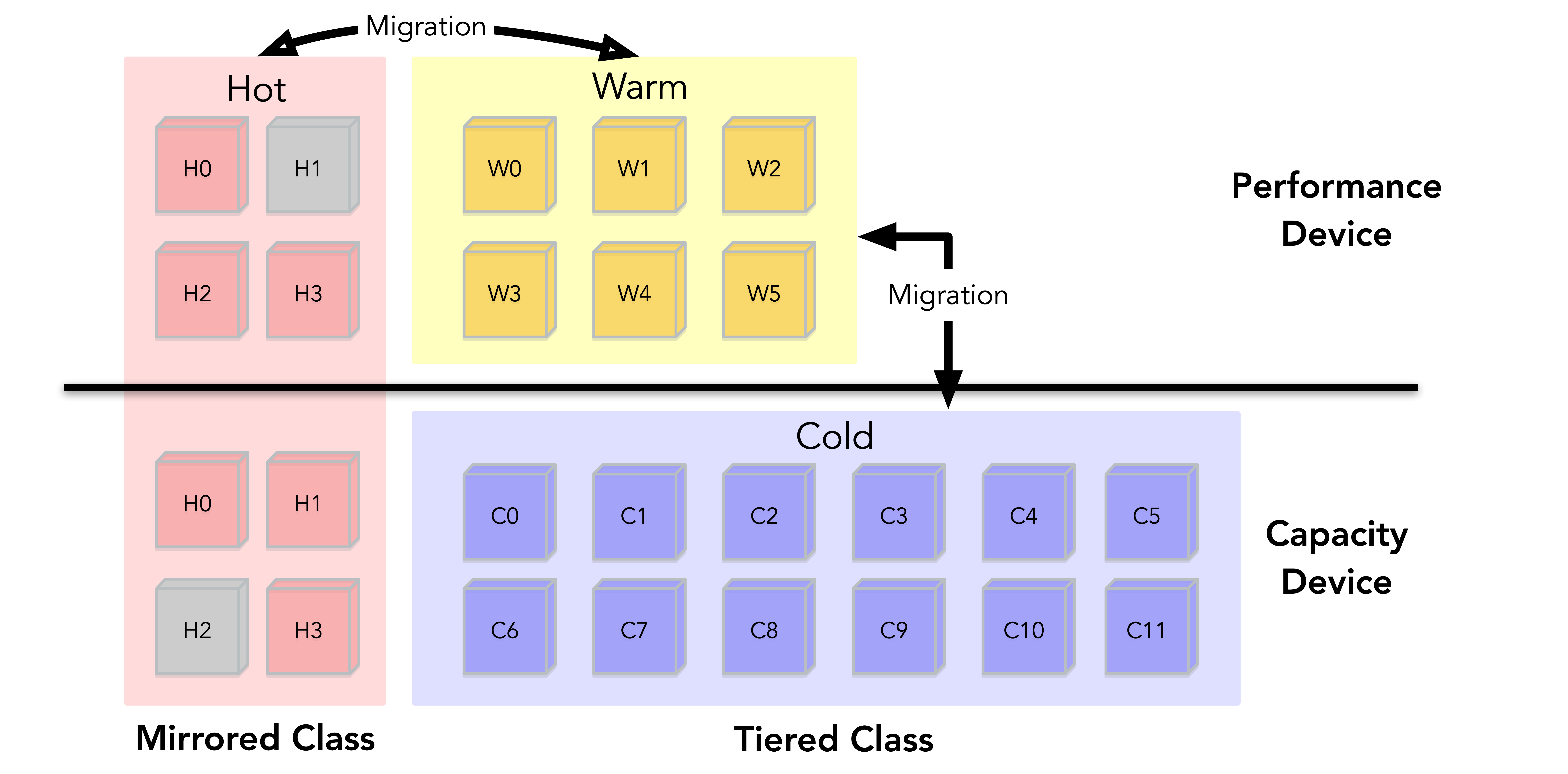}
    \caption{\textbf{\Conceptname Data Layout.} \textit{Data are logically grouped but not physically placed together on the device.}}
    \label{placement}
    \vspace{-0.1in}
\end{figure}
\subsection{Storage vs. Memory Hierarchies}
\label{sec:storage_vs_mem}
The approaches that handle data placement in memory hierarchies~\cite{raybuck2021hemem,vuppalapati2024tiered,lee2023memtis,chou2017batman} versus storage hierarchies~\cite{yang2017autotiering,wilkes1996hp,zheng2019ziggurat} have many similarities.  In both the memory and the storage domain, strategies for data placement optimize for many of the same metrics, such as high performance (bandwidth and latency) and high capacity.  In both domains, solutions share many of the same high-level approaches (e.g., striping, tiering, caching), as well as similar low-level techniques (e.g., attempting to balance load across heterogeneous devices by reacting to observed performance whether for memory~\cite{chou2017batman, vuppalapati2024tiered} or for storage~\cite{wu2021storage}).

However, storage hierarchies have characteristics that require special attention and introduce additional opportunities.
%\begin{enumerate}[noitemsep, topsep=0pt]
\noindent {\bf Larger Capacity and Datasets:} Storage devices manage significantly larger datasets than memory.  Thus, if migration is used as the primary technique for balancing load across tiers, more data will need to be migrated leading to larger convergence times when adjusting to workload changes.

\noindent {\bf Limited Write Bandwidth:} Storage devices have lower write bandwidth than memory; combined with the previous point, this further increases convergence time if migration is used to adjust to dynamic workloads.

\noindent {\bf Read/Write Interference:} Some storage devices~\cite{wu2019contract, wu2022nyxcache} have complex performance characteristics such that write operations significantly impact the performance of read operations.  Migrating data in the background introduces write traffic that can significantly degrade foreground performance.

\noindent {\bf Device Endurance:} Frequent migrations accelerate device wear, reducing the lifespan of storage devices such as SSDs~\cite{mcallister2021kangaroo, mcallister2024fairywren}.

\noindent {\bf Software-based Indirection:}  Since storage devices have slower access time (10 - 500~$\mu s$) than memory (50 - 100~ns), software is used to determine the location of blocks on different devices instead of hardware.  As a result, techniques such as dynamic and selective mirroring are more straightforward to implement for the storage stack.   
%\end{enumerate}

Thus, while dynamic tiering with data migration has been shown to work well for heterogeneous memory hierarchies~\cite{vuppalapati2024tiered, raybuck2021hemem}, heterogeneous storage hierarchies need additional techniques that do not rely as heavily on migration.

\section{\conceptnamefull}
\label{sec:approach}
We present \textit{\conceptnamefull} (\conceptname), a combination of mirroring and tiering that enables flexible load balancing across tiers by mirroring a small amount of data. \conceptname is based on three design goals: 

\noindent
\textbf{Maximized Bandwidth Utilization.} \conceptname should dynamically balance load across all storage tiers based on workload.

\noindent
\textbf{Quick Response to Dynamic Workloads.} \conceptname should dynamically and quickly adapt to varying workloads, including fluctuations in load and working set size.

\noindent
\textbf{Independence from Device and Workload Characteristics.} \conceptname should operate effectively across any storage hierarchy without requiring prior knowledge of device characteristics or workload patterns.

% read/write mixed workload

\if 0
Classic tiering places data in either the performance or capacity device, but because data accesses must be served from the data's current location, classic tiering has no control over load distribution across devices without migrating data.
\fi

\subsection{Basic Design}
The key idea of \conceptname is to combine the load-balancing advantages of mirroring with the capacity benefits of classic tiering.   \conceptname uses a hybrid data layout with two classes: the mirrored class and the tiered class (Figure~\ref{placement}). Data in the mirrored class is replicated across two devices to enable fast load balancing; data in the tiered class is stored as a single copy to maximize the space efficiency. In \conceptname, \textit{hot} data, which is accessed most frequently, is moved to the mirrored class and the remaining data is stored in the tiered class. Within the tiered class, \textit{warm} data is migrated to the performance device; the remaining \textit{cold} data is migrated to the capacity device. 

As desired, the performance device stores the most-accessed data: \textit{hot} data in the mirrored class and \textit{warm} data in the tiered class.  Under light workloads, this layout maximizes the likelihood that a request can be served from the performance device. Under intense workloads, \conceptname\ dynamically increases the size of the mirrored class and routes some requests in the mirrored class to the capacity copy to balance the load across both devices.

% which is accessed frequently but colder than the hot data 
% \input{simulation}

\begin{figure}[t!]
    \centering
    \includegraphics[width=0.8\columnwidth]{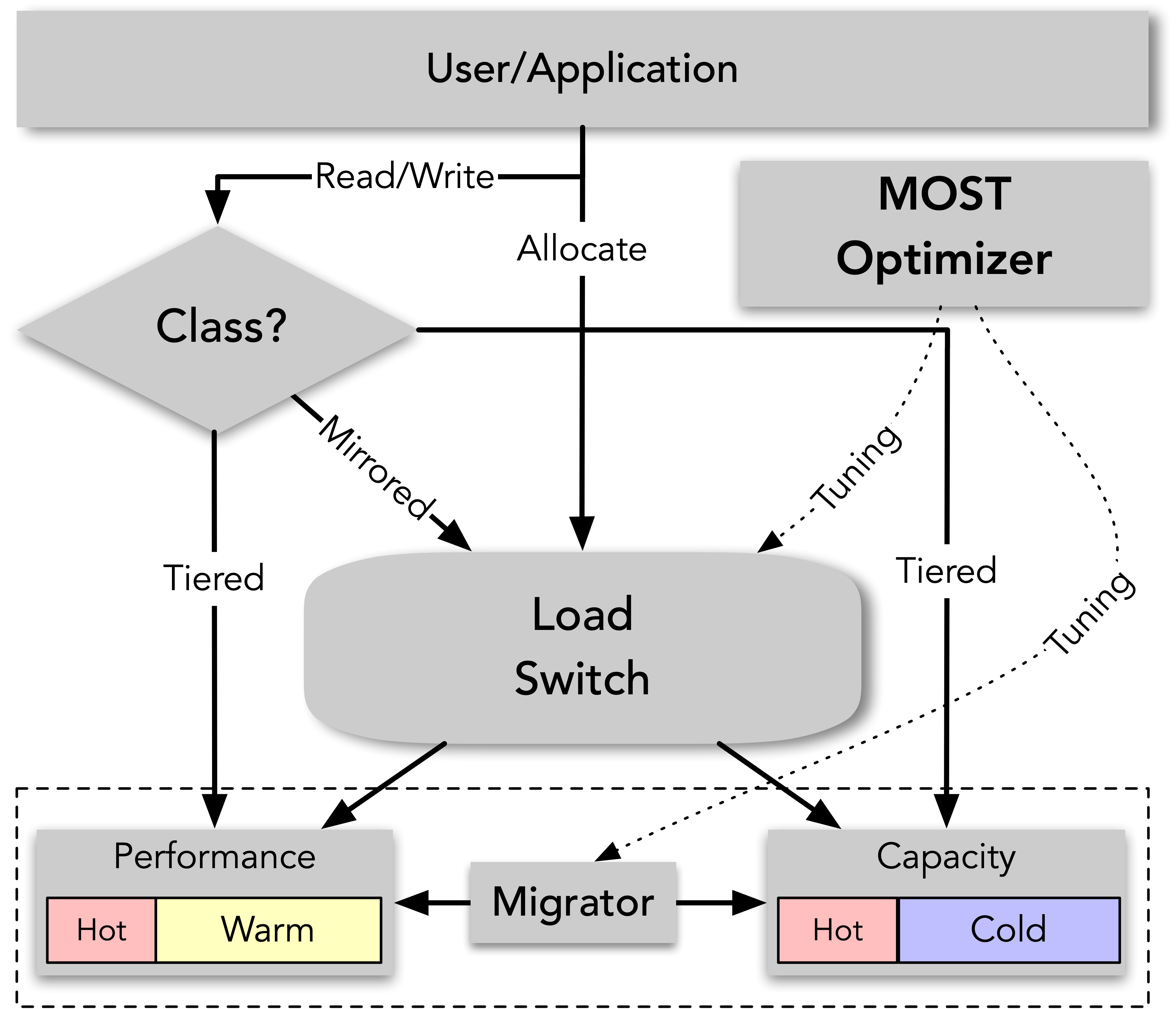}
    \caption{\textbf{\Conceptname Architecture.}}
    \label{architecture}
    \vspace{-0.15in}
\end{figure}

\begin{algorithm}[t!]
    \caption{\textbf{\Conceptname Optimizer}}
    \begin{algorithmic}[1]
    \While{true}
        \State sleep(\textit{tuningInterval}); Measure end-to-end latency.
        \If{$L_{P}$ > (1 + $\theta$) * $L_{C}$}
            \If{\textit{offloadRatio} \( = \) \textit{offloadRatioMax}}
                \If{\textit{Mirrored class is not maximized}}
                    \State \textit{Enlarge the mirrored class}
                \Else
                     \State \textit{Improve hotness of the mirrored class}
                \EndIf
                \State \textit{only migrate to capacity device}
            \Else { \textit{offloadRatio} \( \mathrel{\text{+=}} \) \textit{ratioStep}}
                
            \EndIf
        \ElsIf{$L_{P}$ < (1 - $\theta$) * $L_{C}$ }
            \If{\textit{offloadRatio} \( = \) 0}
                % \State \textit{mirroredClassSize} -= \textit{sizingStep}
                \State  \textit{only migrate to performance device}
            \Else { \textit{offloadRatio} \( \mathrel{\text{-=}} \) \textit{ratioStep}}
            \EndIf
        \Else { \textit{stop all migration}}
        \EndIf
    \EndWhile
    \end{algorithmic}
    \label{most_algorithm}
    % \vspace{-0.2in}
    % \vspace{-1em} % Shrinks space inside the float at the end.
\end{algorithm}

\subsection{Design Details}

Figure~\ref{architecture} shows the basic architecture of \conceptname.  Blocks are placed in the mirrored or tiered class depending on their frequency of access: the goal is to place the hottest data in the mirrored class, warm data in the tiered class on the performance device, and cold data in the tiered class on the capacity device. When handling a request, \conceptname\ first determines if the request is to the tiered or mirrored class.  Requests to tiered data are straightforward since the data resides strictly in either the performance or capacity device; those reads and writes are simply forwarded to the appropriate location.  Requests to mirrored data involve more complexity: a load switch (or balancer) sends some percentage of these requests to each device, where the percentage depends upon the current workload performance and whether the request is a read or write.  Other components of the architecture calculate this percentage (i.e., Optimizer) and migrate data between the classes and devices (i.e., Migrator).  

We address key questions of \conceptname's design.
%\begin{itemize}[noitemsep, topsep=0pt]
For reads to the hottest data, which data copy in the mirrored class should be read to balance traffic across devices? 
On which device should newly written data be allocated?
What data should be migrated between mirrored and tiered classes, and between devices?
For writes to mirrored data, which copy should be updated?  At what granularity should clean copies be tracked?  When should copies be cleaned?
How to protect the tail latency of mirrored data?
%\end{itemize}

\subsubsection{Balancing Read Traffic in the Mirrored Class}
For simplicity, we begin by focusing on reads that are to the hottest data and thus in the mirrored class. We later show how \conceptname\ allocates data, moves data to and from the mirrored class, and handles writes.  The full algorithm is shown above. 
%(Alg. \ref{most_algorithm}). 

At a high level, reads to mirrored data should be routed to the device that delivers the best performance. To achieve this, \conceptname employs probability-based routing to determine which copy to read~\cite{wu2021storage}. Specifically, \conceptname routes an incoming read with probability \textit{offloadRatio} to the capacity device, and otherwise, to the performance device. \textit{OffloadRatio} is the key variable controlling load distribution, as well as write traffic and allocation. \textit{OffloadRatio} is tuned by a simple yet efficient feedback-driven dynamic optimization algorithm within the Optimizer; \textit{offloadRatio} is adjusted such that the measured end-to-end latency of the devices is equalized. 

Specifically, assuming $L_P$ is the latency of the performance device and $L_C$ is the latency of the capacity device, then

\begin{itemize}[noitemsep, topsep=0pt]
    \item When $L_{P} < L_{C}$, accessing the performance device is faster than accessing the capacity device, and so more traffic should be routed to the performance device (see Lines 11–14); thus, \textit{offloadRatio} is decreased.  
    \item When $L_{P} > L_{C}$, accessing the capacity device is faster than accessing the performance device, and so more traffic should be routed to the capacity device (see Lines 3–10); thus, \textit{offloadRatio} is increased.  
    \item When $L_{P} = L_{C}$, the two access latencies are approximately  equal, and therefore, no further action is required (Line 15).
\end{itemize}

\conceptname efficiently adapts to both high-load and low-load scenarios. Under low load, the latency of accessing the performance device is lower than that of the capacity device, and so traffic is routed to the performance device; in these conditions, \conceptname operates like classic tiering, directing as many requests as possible to the performance device.  In contrast, under high load, all hot and warm data reside on the performance device, causing its access latency to increase—due to queuing and internal resource contention—and eventually exceed that of the capacity device; consequently, \conceptname offloads requests to the capacity device which improves throughput.  Under high load, MOST offloads traffic from the performance device to the capacity device until the end-to-end latency between the two devices is equalized.

%%%%%%%%%%%%%%%%%%%%%%%%%%%%%%%%%%%%%%%%%%%%%%%%%%%%%%%%%%%%%%%%%%%%%%%%%%%%%%%%%%%%%%%%%%%%%%%%%%%%%%%%%%%%

%\noindent \textbf{Dynamic Write Allocation.}
\subsubsection{Dynamic Write Allocation}
When a data block is first written, \conceptname\ allocates its location. MOST divides both mirrored and tiered storage into fixed-sized segments (e.g., 2MB) and maintains in-memory data structures to track the metadata for the segment. All data are initially allocated into the tiering class.  In classic tiering, the allocation policy is load-unaware, meaning that it always allocates newly-created blocks on the performance device, even when that device is saturated. \Conceptname introduces a probability-based write allocation policy: similar to load balancing within the mirrored class, the newly allocated data is placed on the capacity device with \textit{offloadRatio} probability.  As a result, when the performance device is under high load (higher latency), more data is allocated on the capacity device; when the performance device is under light load (lower latency), all data is allocated on the performance device, as desired and as in classic tiering~\cite{raybuck2021hemem,vuppalapati2024tiered}.  After a block has been allocated to the tiered class, it may later be promoted to the mirrored class, as described next.

\subsubsection{Mirror-Class Migration} 
To meet the basic goal of placing the hottest data in the mirrored class, \conceptname\ migrates data between classes such that it minimizes the amount of movement between devices.   To identify hot/warm/cold data, \conceptname tracks read and write counters for each segment, similar to HeMem~\cite{raybuck2021hemem}; thus, hotness is based on access frequency.

%if there is spare space in the system, 
When the amount of mirrored data is insufficient to effectively balance traffic (Line 4), \conceptname increases the size of the mirrored class up to a configured maximum. Our experiments (\shortref{sec:eval}) show that devoting 20\% of the total capacity to the mirrored class is sufficient for our workloads.   To migrate data into the mirrored class, \conceptname selects the hottest segment from the tiered class on the performance device; this hottest data is simply duplicated onto the mirrored class on the capacity device.   If the mirrored class reaches its maximum size and the hottest segment in the tiered class has a higher access frequency than the coldest segment in the mirrored class, \conceptname swaps those segments.  Space within the mirrored class is reclaimed when the available system capacity drops below a predefined watermark (2.5\% of the total capacity). During reclamation, \conceptname reclaims the coldest segment in the mirrored class: if a valid copy of this segment exists on the performance device, the capacity copy is discarded; otherwise, the copy on the performance device is discarded.  

% To minimize migration, MOST only promotes data on the performance device to the mirrored class.  When space is available in the mirrored class, the hottest data blocks on the performance device in the tiered class are promoted until the mirrored class is full (or no data remains).

\noindent \textbf{Migration Regulation.}
In classic tiering, hot data is always migrated to the performance device and cold data to the capacity device; however, with \conceptname\ and its mirrored class, hot segments may be migrated to the capacity device.  \conceptname dynamically manages this bidirectional migration, adhering to a principle of migrating exclusively {\it away} from the device experiencing higher end-to-end latency. Precisely, when the performance device has higher latency, migration to the performance device is stopped and migration to the capacity device is enabled; when the capacity device exhibits higher latency, the opposite is performed; if the latencies of both devices are approximately equal, all migration is stopped.

%%%%%%%%%%%%%%%%%%%%%%%%%%%%%%%%%%%%%%%%%%%%%%%%%%%%%%%%%%%%%%%%%%%%%%%%%%%%%%%%%%%%%%%%%%%%%%%%%%%%%%%%%%%%

\subsubsection{Balancing Writes and Tracking Clean Subpages}

Our previous discussion focused on reads; writes need special consideration.  Writes to the tiered class remain straightforward since there is only one copy to update.  However, writes to the mirrored class have a new property: if the write is performed to both copies, then the system is not performing any write load-balancing; however, if the write updates only one copy, then future reads must be directed only to that clean, valid copy.  In order to perform load-balancing of writes, MOST updates only one copy  and carefully tracks which portions of each segment are valid.

% On writes, \conceptname updates only a single copy with write load-balancing, dynamic write allocation, and selective periodic cleaning.  For writes to the mirrored class, a fixed percentage is routed to each device.  Future reads and writes to this data, which has only one clean copy, are straightforward: they are served from the device holding the clean copy.  

Writes to the mirrored class are balanced as follows.   If both copies of the mirrored data are valid, the write request is probabilistically sent to either the capacity or performance device based on \textit{offloadRatio}.  However, if only one copy in the mirror is valid, a naive implementation based only on segments must direct later write traffic only to the valid segment (or overwrite the entire invalid segment, making it valid).  

\if 0
either overwrite the entire segment or 
implementation with the write request must be sent to the valid copy to avoid a read-modify-write operation, which would trigger an additional read request unless the write request overwrites the entire 2MB segment. Consequently, if the workload has a write request size smaller than 2MB, these write operations cannot be dynamically routed but only sent to one copy of the mirrored data.
\fi

%\changed{\noindent \textbf{Subpage Management of Mirrored Data.} Modern applications typically issue storage requests aligned to the native sector granularity of modern storage devices\cite{}. 

\noindent \textbf{Mirrored Data Subpages.} To enable better load balancing of writes, \Conceptname manages segments in the mirrored class at a finer granularity: an invalid bit and a location bit are tracked for each subpage corresponding to the device's unit of access (e.g., 4KB).  Each subpage exists in one of three states: clean (both copies are valid), invalid on the performance device, or invalid on the capacity device.  Thus, given an aligned subpage write to the mirrored class, \conceptname can route the write to either device without having to update the entire segment; that is, a 4KB-aligned write can be load balanced through simple routing, similar to reads. Subpages in the mirrored class slightly increase metadata overhead (2 bits of metadata), but since the mirrored class is relatively small, the overall overhead remains minimal; for instance, in a 2TB hierarchy in the extreme case where all performance device data is mirrored (50\% mirroring), the metadata overhead is only 128MB, which is negligible.

\noindent \textbf{Selective Cleaning in the Mirror Class.} Data blocks with only one valid copy are selectively cleaned by a background thread. This cleaning thread selects blocks with a large \textit{rewrite distance}, the average number of reads between two writes for a given block. We have observed that when a block has a small \textit{rewrite distance}, it is likely to be rewritten soon, making cleaning ineffectual.

\subsubsection{Tail Latency Protection}
The description thus far has focused on maximizing total bandwidth from the performance and capacity devices; however, \conceptname allows users to protect the tail latency of mirrored (hot) data by setting a maximum \textit{offloadRatio} which limits the traffic offloaded to the capacity device in the mirrored class when the capacity device shows significantly worse tail latency than the performance device.

\begin{figure}[t!]
    \centering
    \includegraphics[width=\columnwidth]{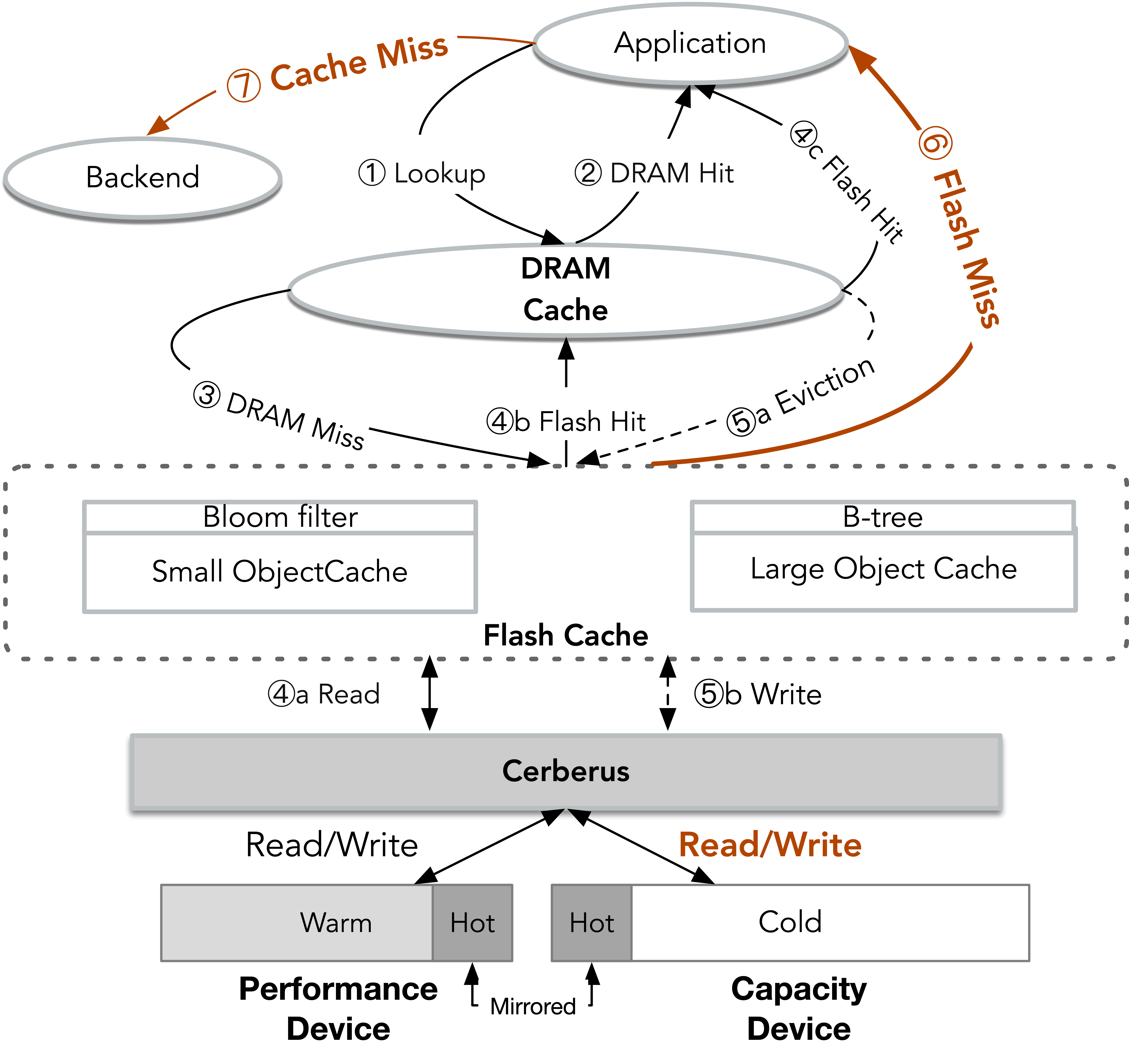}
    \caption{\textbf{CacheLib Architecture.} This figure shows CacheLib’s architecture and the lookup workflow. A lookup first checks the DRAM cache (\protect\encircle{\makebox[0.8em][c]{1}}) and immediately returns the object on a hit (\protect\encircle{\makebox[0.8em][c]{2}}). On a miss, it checks the flash cache (\protect\encircle{\makebox[0.8em][c]{3}}), issuing a read to the underlying storage devices (\protect\encircle{\makebox[0.8em][c]{4a}}); if the object is found, it returns the result (\protect\encircle{\makebox[0.8em][c]{4b}}). A flash cache hit promotes the item to DRAM and may evict an existing DRAM entry (\protect\encircle{\makebox[0.8em][c]{5a}}, \protect\encircle{\makebox[0.8em][c]{5b}}). A miss in the flash cache (\protect\encircle{\makebox[0.8em][c]{6}}) leads to a back-end access (\protect\encircle{\makebox[0.8em][c]{7}}).}
    \label{fig:cachelib}
    \vspace{-0.1in}
\end{figure}

\subsection{Implementation: Cerberus}
\label{sec:implementation}

% \textit{The circled numbers refer to steps described in Section \ref{sec:implementation}.}

% The DRAM layer serves as a cache for the flash SSD. 

We introduce Cerberus, a user-level storage management system that integrates \conceptname into CacheLib.  CacheLib is a generic flash cache library used extensively in data center infrastructures~\cite{berg2020cachelib,netflix,cdnSong,mcallister2021kangaroo,flashshied,tang2015ripq}. As shown in Figure \ref{fig:cachelib}, CacheLib consists of three layers: a DRAM cache layer, a flash cache layer, and a storage management layer. CacheLib provides two default flash cache layers: a Large Object Cache (LOC) employs a log with a DRAM index for items 2KB or larger; a Small Object Cache (SOC) stores key-value pairs in a 4KB bucket hash table.  Users can also implement custom flash caches in this flash cache layer (e.g., Kangaroo~\cite{mcallister2021kangaroo} and Fairy-WREN~\cite{mcallister2024fairywren}). 

By default, the storage management layer in CacheLib only provides striping across devices.  Cerberus is our storage management layer between the flash cache engine and the storage hierarchy employing MOST; Cerberus transparently manages the underlying performance and capacity devices, providing a block interface and a large address space.  In addition to MOST, we have implemented other tiering and caching approaches within the CacheLib storage management layer.

%\begin{itemize}[noitemsep, topsep=0pt]
\noindent \textbf{Striping.} The default implementation in CacheLib.

\noindent \textbf{Orthus.} The non-hierarchical caching implementation described and provided by Wu et al.~\cite{wu2021storage} in .6K lines of code (LOC).

\noindent \textbf{HeMem.} Our implementation of the classic tiering and migration algorithm~\cite{raybuck2021hemem} in about .7k LOC.  The original HeMem uses a quantum of 10ms which is appropriate for memory latency, but not storage; we use 200ms, the smallest interval that accurately reacts to our storage device latency.

\noindent \textbf{BATMAN.} Our implementation of BATMAN heterogeneous tiering~\cite{chou2017batman}, requiring about .4K LOC.

\noindent \textbf{Colloid.} Our implementation of Colloid~\cite{vuppalapati2024tiered} is built on HeMem (with a 200ms quantum) in about .4K lines of code.  Since Colloid only balances reads, we implement Colloid+ to incorporate write latency into decisions. Additionally, since Colloid is sensitive to parameters, we implement Colloid++ with $\theta=0.2$ and $\alpha=0.01$, which improves robustness given storage device performance fluctuations. 
%\end{itemize}

% and does not perform well for write-only workloads,

 % and maintains an in-memory data structure to track metadata for each segment
 
\noindent \textbf{Metadata Management.} \Conceptname divides storage into 2MB segments, where each segment requires 76 bytes of metadata as shown in Table \ref{tab:metadata_breakdown}.  Smaller segments would result in larger metadata overhead (e.g., 4KB segments require 512 times more metadata) and degrade performance since they obtain lower device bandwidth (Table \ref{tab:device_stat}). Conversely, larger segments would lead to inefficient utilization of the performance tier, as only a small fraction of subpages might be frequently accessed.   We have found that 2MB segments balance the metadata footprint while minimizing space waste on the performance tier, matching the choice of many other systems~\cite{raybuck2021hemem,vuppalapati2024tiered,lee2023memtis}.  

\begin{table}[t]
\centering
\footnotesize
% \renewcommand{\arraystretch}{0.2} % Reduce row height
% \resizebox{0.9\columnwidth}{!}{ % Make table wider
\begin{tabular}{lc}
\toprule
\textbf{Member Variable} & \textbf{Size (bytes)} \\ 
\midrule
id (uint64\_t)                       & 8 \\
addr[2] (uint64\_t[\,])               & 16 \\
invalid (bitset<512>*)               & 8 \\
location (bitset<512>*)              & 8 \\
clock (uint64\_t)                    & 8 \\
readCounter (uint8\_t)     & 1 \\
writeCounter (uint8\_t)    & 1  \\
rewriteReadCounter (uint64\_t)    & 8  \\
rewriteCounter (uint64\_t)    & 8  \\
flags (uint8\_t)                     & 1 \\
storageClass (enum class)            & 1  \\
mutex (SharedMutex)                    & 8 \\
\midrule
\textbf{Total} & \textbf{76} \\ 
\bottomrule
\end{tabular}
% }
\caption{\textbf{In-Memory Metadata per Segment.}}
\label{tab:metadata_breakdown}
\vspace{-0.1in}
\end{table}

\noindent \textbf{Implementation Details.} The MOST optimizer runs on a single pinned thread during each 200ms interval.  At each interval, the optimizer estimates the access latency of each device by comparing counters from the Linux block-layer~\cite{linux2023block} to measurements from the previous interval. Similar to prior systems~\cite{vuppalapati2024tiered}, we apply Exponentially Weighted Moving Average (EWMA) to measured latency to smooth out short-term fluctuations and maintain long-term stability. The optimizer leverages this smoothed latency signal to guide migration decisions and adjust routing accordingly. We set $\theta = 0.05$, a commonly used tolerance in tuning-based systems\cite{vuppalapati2024tiered}, to treat values as approximately equal, striking a balance between stability and responsiveness. MOST demonstrates robust performance across diverse workloads without requiring fine-tuning, indicating low sensitivity to the specific choice of $\theta$. We also adopt \textit{ratioStep} = 0.02, following guidance from similar systems (e.g., Orthus~\cite{wu2021storage}), which works well across different workloads. Cerberus introduces approximately 1.5k LOC to CacheLib, leveraging and extending the core HeMem tiering logic.

\if 0
Cerberus: 1540
HeMem: 720
Caching: 656
BATMAN: 378
\fi

% How the latency is measured in detail with which field is using. How accurately it is compared to the real scenario. 
% How do we handle the read/write latency. Do we mixed together? 
% How do we estimate the unloaded latency and how do we get the average request size through diskstats. 
% How the average latency of read and write is calculated (weighted avearge latency of read and write).

% All experiments are conducted on a static storage configuration where both devices are local.
\begin{figure*}[t]
  \centering
  \begin{minipage}[b]{.24\linewidth}
    \centering
    \includegraphics[width=\linewidth]{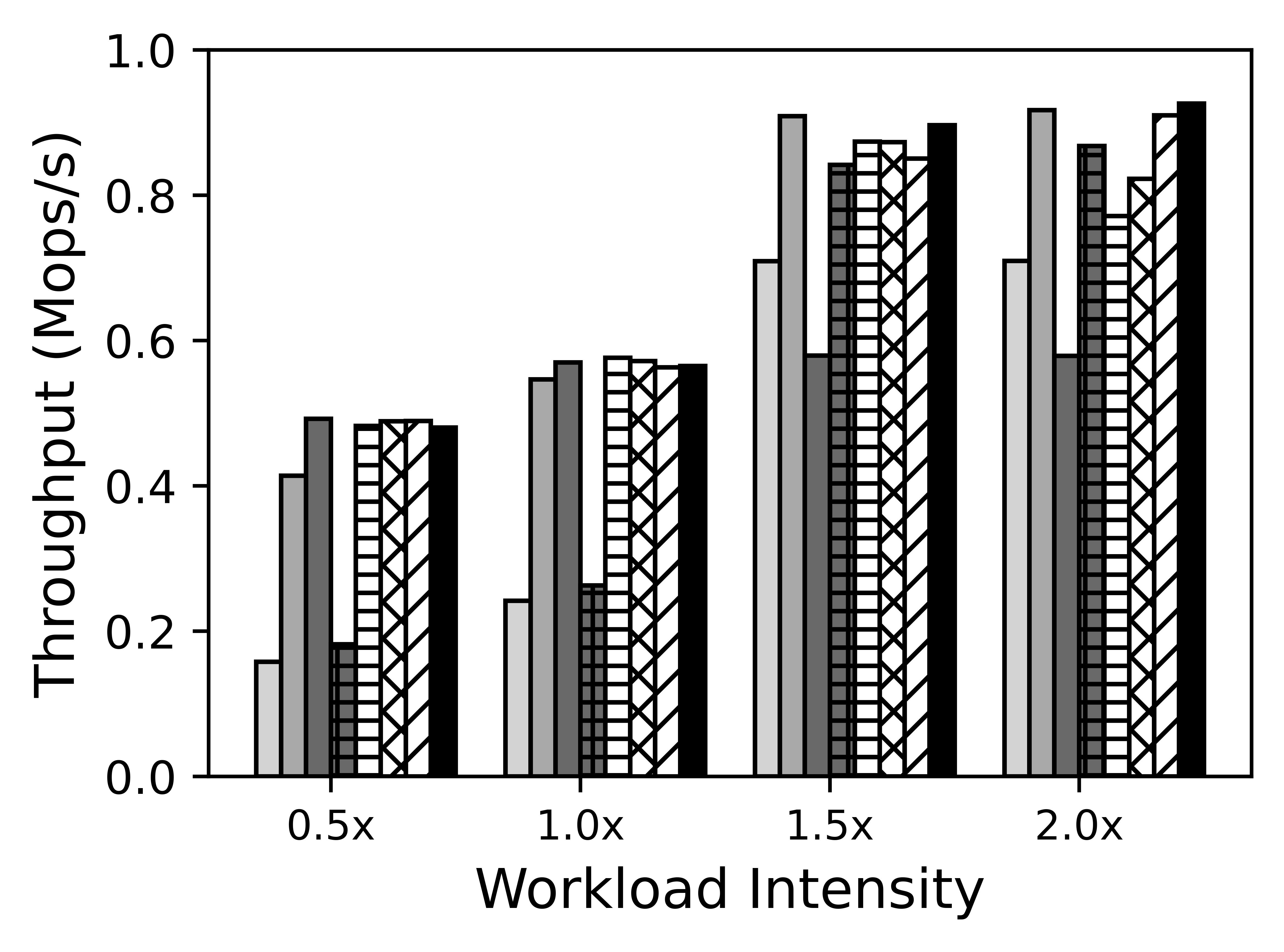}
     \subcaption{\textbf{Random Read-only}}
    \label{fig:static_readonly}
  \end{minipage}%
  \begin{minipage}[b]{.24\linewidth}
    \centering
    \includegraphics[width=\linewidth]{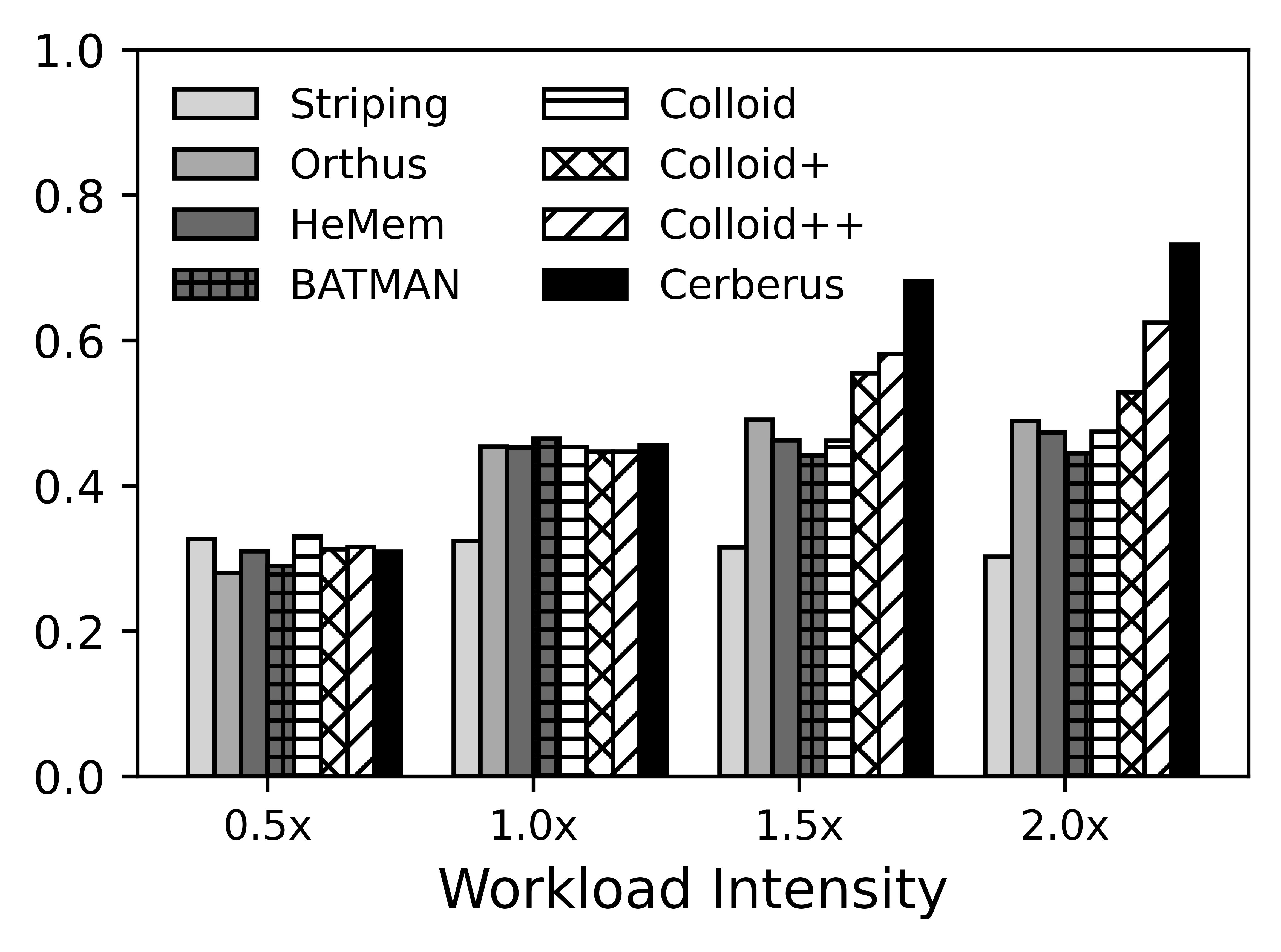}
    \subcaption{\textbf{Random Write-only}}
    \label{fig:static_writeonly}
  \end{minipage}
  \begin{minipage}[b]{.24\linewidth}
    \centering
    \includegraphics[width=\linewidth]{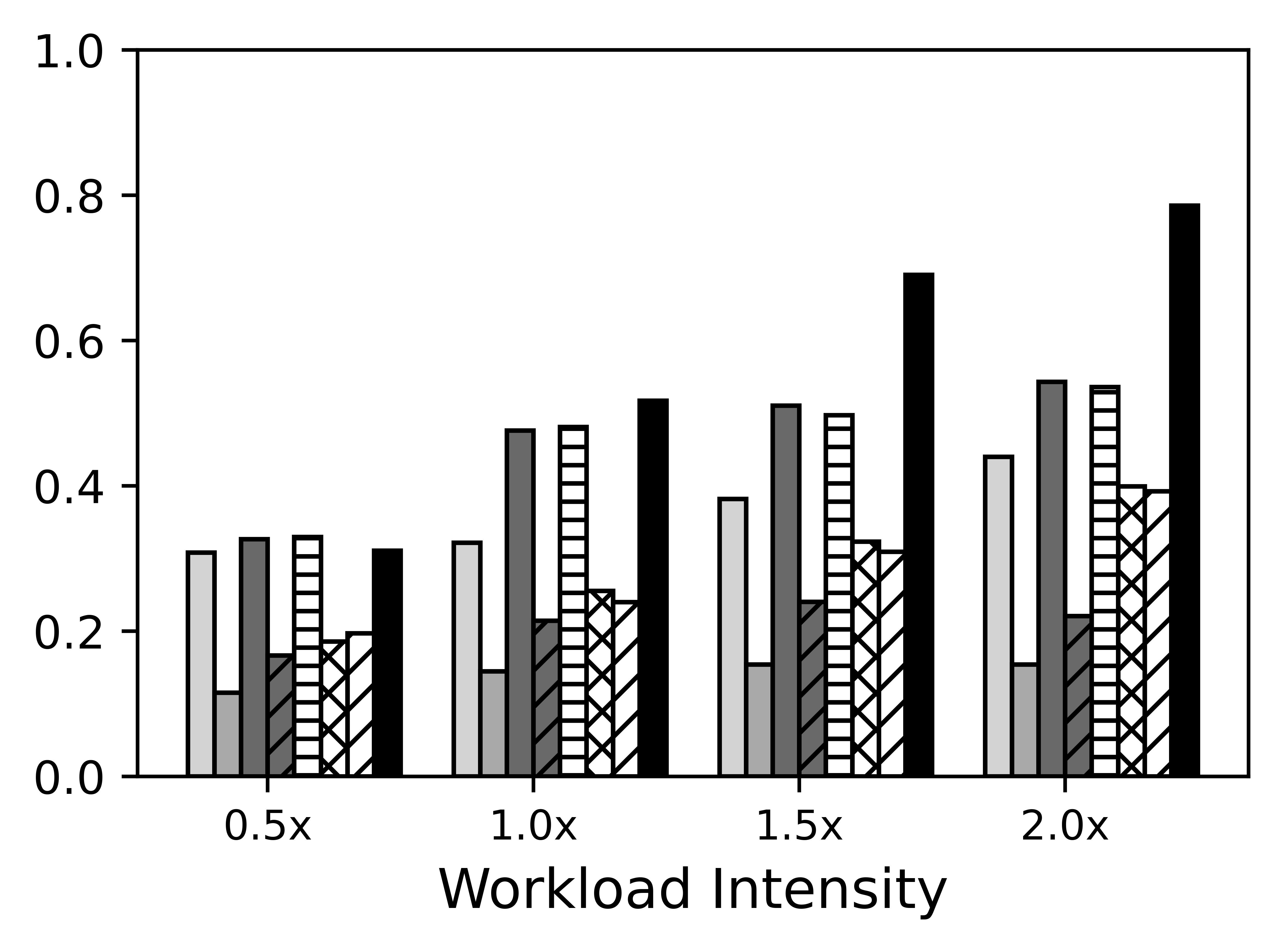}
    \subcaption{\textbf{Sequential Writes}}
    \label{fig:static_sequential}
  \end{minipage}
  \begin{minipage}[b]{.24\linewidth}
    \centering
    \includegraphics[width=\linewidth]{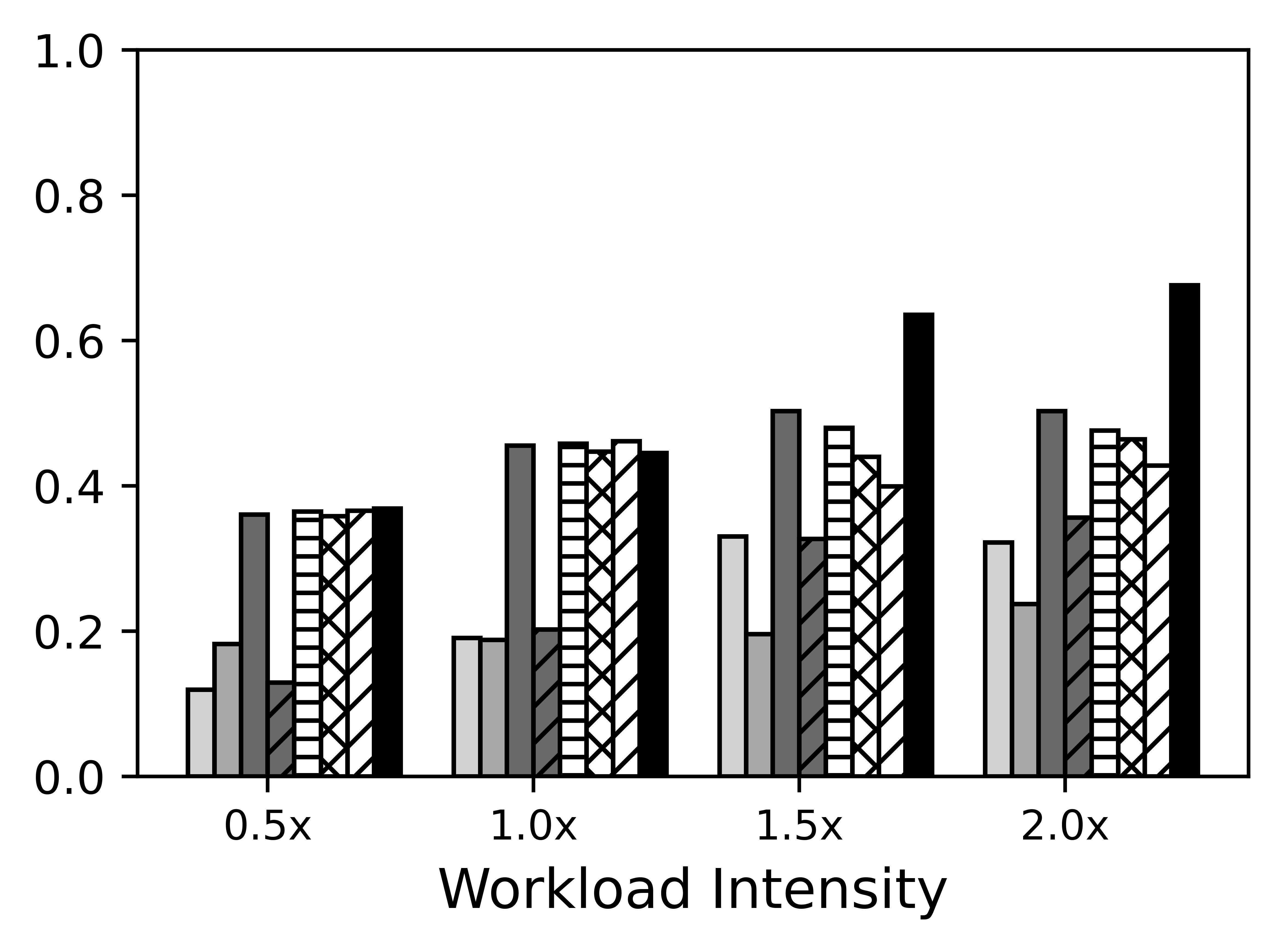}
    \subcaption{\textbf{Read Latest}}
    \label{fig:static_readrecent}
  \end{minipage}
  \caption{\textbf{Static Workload.} \textit{The workload is running on the Optane/NVMe hierarchy with 750GB working set. In (a), under intensity 2.0$\times$, Colloid, Colloid+, Colloid++, Cerberus migrate 134GB, 122GB, 62GB, 50GB of data in total. In (b), under intensity 2.0$\times$, Colloid, Colloid+, Colloid++, Cerberus migrate 0GB, 93GB, 80GB, 65GB of data in total.}}
  \label{fig:micro_static}
  \vspace{-0.1in}
\end{figure*}
\section{Evaluation}
\label{sec:eval}
We evaluate Cerberus to answer the following questions.
\begin{itemize}[noitemsep, topsep=0pt]
    \item How does Cerberus compare to striping, caching (Orthus), and tiering (HeMem, BATMAN, and Colloid) for static workloads with different intensity levels and patterns? (\shortref{sec:micro_static})
    \item How quickly does Cerberus adapt to dynamic workloads? How does the number of writes performed by MOST compare to migration-based tiering approaches with load-balancing capabilities? (\shortref{sec:micro_dynamic})
    \item How effective are Cerberus's techniques for mirror class sizing, subpage management, and selective cleaning? (\shortref{sec:in_depth})
    \item How does CacheLib with Cerberus perform on static workloads? How does it perform on dynamic workloads? (\shortref{sec:cachelib})
\end{itemize}

\noindent\textbf{Storage Configurations.}  We evaluate two storage hierarchies: Optane (performance) / NVMe (capacity), and NVMe (performance) / SATA (capacity).  The latency and bandwidth of these three storage devices (750GB Intel Optane SSD P4800X, 1TB Samsung 960 NVMe SSD, 1TB Samsung 870 SATA SSD) are shown in Table \ref{tab:device_stat}.  The server has a 40-core Intel Xeon Gold 5218R CPU @ 2.1GHz (Ubuntu 20.04) and 64GB DRAM.

\subsection{Comparison to Previous Approaches}
\label{sec:micro_static}
We begin with a high-level comparison of Cerberus to other approaches for handling heterogeneous devices: striping, Orthus (caching), HeMem (classic tiering), BATMAN (with a static ratio matching the read bandwidth of the devices), and three versions of Colloid (a state-of-the-art tiering approach).  To focus on the basic performance of these algorithms, we isolate the storage management layer from CacheLib and exercise that layer with controlled workloads; we begin with a static micro-benchmark in which multiple threads perform synchronous random reads and writes, given a skewed access pattern where a 20\% hotset is accessed with 90\% probability.  Figure~\ref{fig:micro_static} compares steady-state throughput for read-only, write-only, sequential write, and read-latest workloads with varying intensities; 1.0× represents the minimum load at which the bandwidth of the performance device is saturated.
We show that on static workloads Cerberus performs as well as or better than all other approaches.

\noindent \textbf{Random Read-only.}  Even on this simple static workload, the other approaches do not match Cerberus. 
Striping delivers suboptimal performance, since it is bottlenecked by the slower device. Orthus delivers similar throughput to Cerberus, but Orthus achieves this by mirroring 690GB of data compared to only 50GB in Cerberus. HeMem (classic tiering) reaches its performance limit when the load intensity is 1.0$\times$ because HeMem does not offload traffic to the capacity device once the performance device's bandwidth is saturated; consequently, further load increases do not improve throughput.   For BATMAN, no static allocation ratio works well for all load levels: BATMAN performs relatively well under high load because its allocation ratio is configured to match this bandwidth ratio for the hierarchy; however, at low load, this ratio causes traffic to be sent to the capacity device, reducing throughput.  Colloid experiences significant performance degradation at intensity 2.0$\times$ due to migrations triggered by latency spikes arising from background activity; \textit{Colloid++} improves this throughput, showing that Colloid is sensitive to parameter choice. As a result, Colloid and Colloid++ incur 2.68× and 1.24× more migration traffic than Cerberus (as shown in the caption). In summary, Cerberus delivers high throughput at both low and high load and also reduces device writes compared to migration-based approaches that attempt to balance the load.

\noindent \textbf{Random Write-only.}  We focus on those results that are noticeably different for writes versus reads.  Orthus has a static write-back policy, and so does not balance write traffic or scale under high load.   BATMAN no longer performs well at high load because a different allocation ratio is required to match the performance of write traffic versus read traffic across the two devices.   Colloid does not perform write balancing and therefore exhibits performance similar to HeMem; \textit{Colloid+} balances write traffic but suffers from migration overhead triggered by latency spikes, resulting in suboptimal throughput; \textit{Colloid++} demonstrates improved performance, but still incurs migration overhead.  For write-intensive workloads, Cerberus performs significantly better than other approaches, since it balances writes in the mirrored class and is robust to latency spikes.

\begin{figure*}[t]
  \centering
  \begin{minipage}[b]{.33\textwidth}
    \centering
    \includegraphics[width=\linewidth]{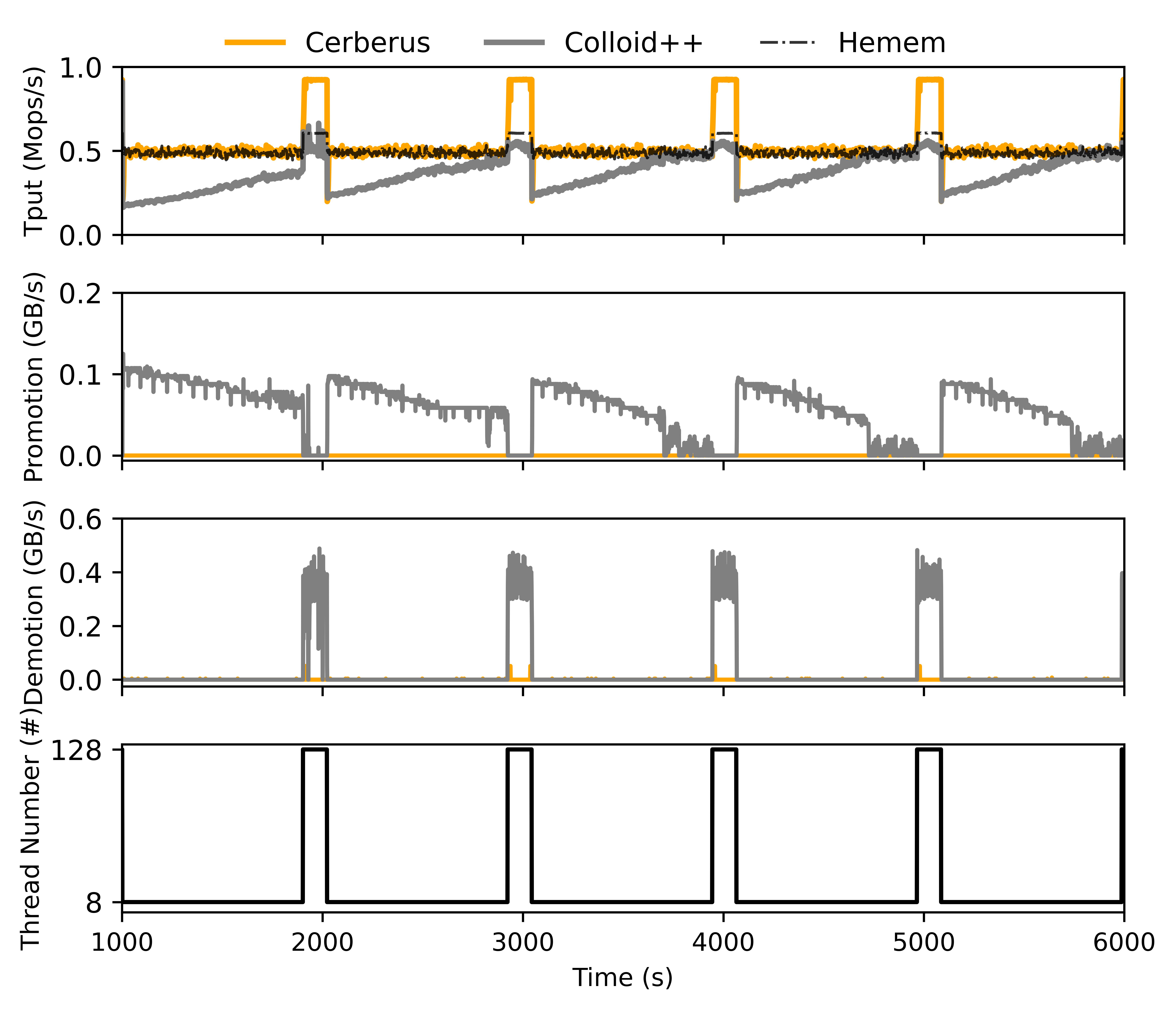}
     \subcaption{\textbf{Read-only}}
    \label{fig:dynamic_readonly}
  \end{minipage}%
  \begin{minipage}[b]{.33\textwidth}
    \centering
    \includegraphics[width=\linewidth]{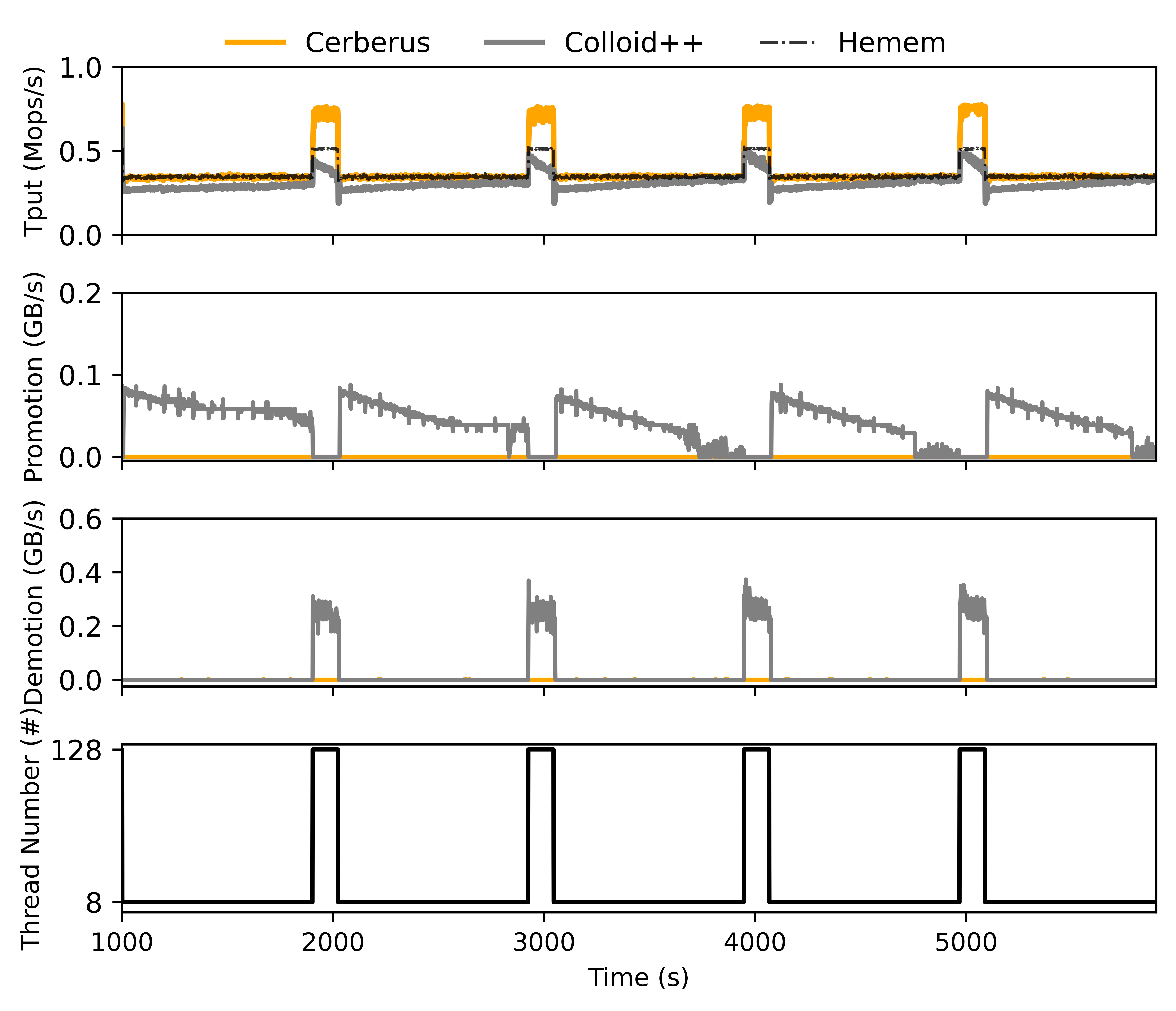}
    \subcaption{\textbf{Write-only}}
    \label{fig:dynamic_writeonly}
  \end{minipage}
   \begin{minipage}[b]{.33\textwidth}
    \centering
    \includegraphics[width=\linewidth]{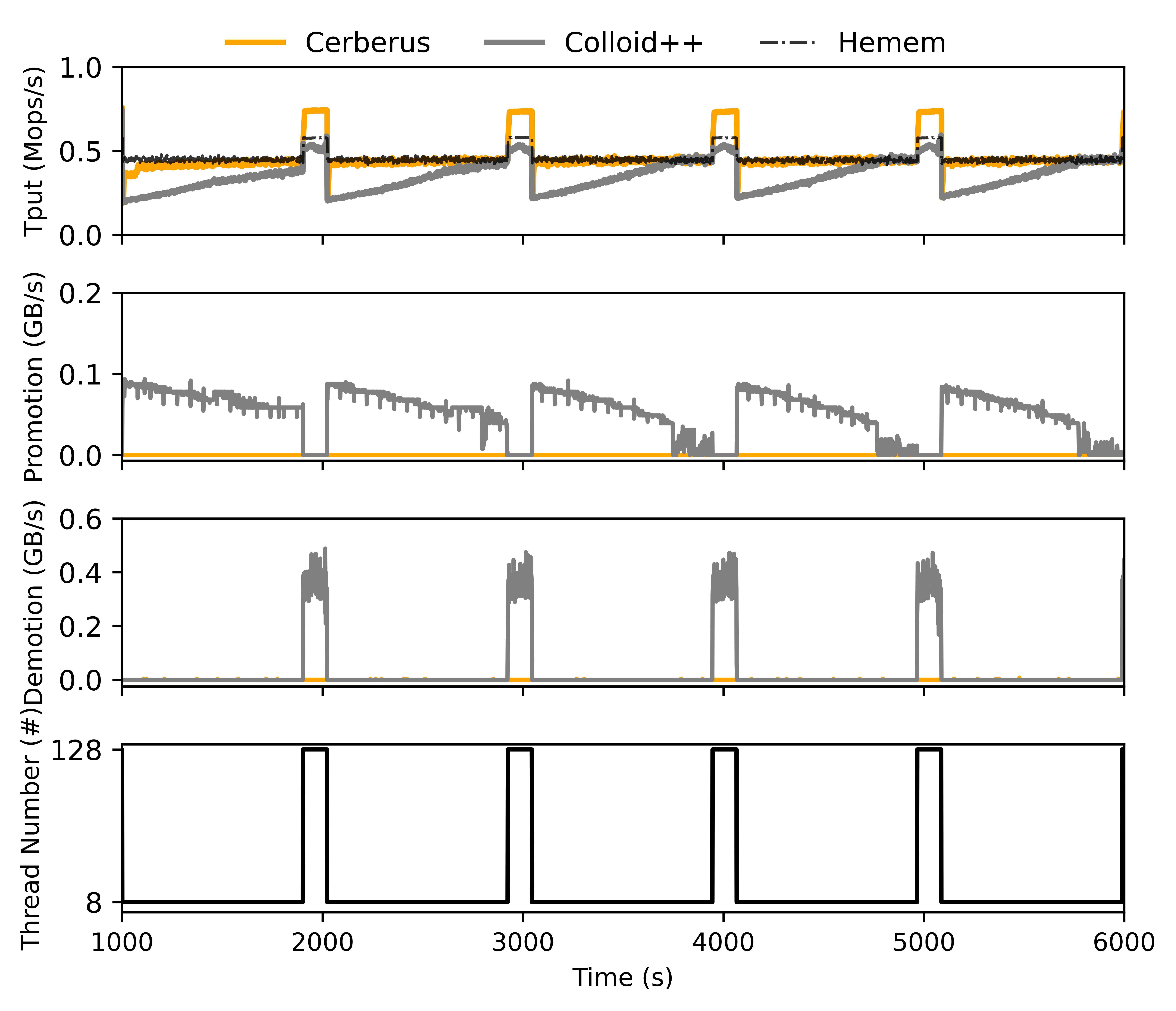}
     \subcaption{\textbf{RW-mixed}}
    \label{fig:dynamic_rwmixed_micro}
  \end{minipage}
  \caption{\textbf{Dynamic Bursty Workload.} \textit{Running on Optane/NVMe hierarchy with 1.2TB working set and same skewness as \shortref{sec:micro_static}. Colloid++ migrates 282GB, 214GB, 260GB of data to the performance device and 262GB, 186GB, 239GB of data to the capacity device under workload (a), (b) and (c), respectively. In comparison, Cerberus migrates 87GB, 107GB and 64GB of data to the capacity device.} }
  \label{fig:micro_dynamic}
  \vspace{-0.1in}
\end{figure*}

\noindent \textbf{Sequential Write.} Sequential writes emulate popular applications with log-structured data (e.g., flash caches, file systems, and databases). As with random writes, Orthus cannot efficiently balance write traffic.   Colloid behaves similarly to HeMem because it doesn't take write latency into account when balancing.   Interestingly, \textit{Colloid+} and \textit{Colloid++} perform worse due to additional interfering migrations when balancing write latency; in a sequential workload, demoted blocks are not re-accessed, which makes these migrations ineffective.  As desired, Cerberus dynamically allocates new writes proportionally across devices: when the performance device becomes saturated, writes are allocated directly on the capacity device.

\noindent \textbf{Read Latest.} This workload uses a 50\% write ratio, where 20\% of the newly-written blocks have a 90\% probability of being read.   Colloid performs worse than HeMem at intensities higher than 1.0× due to migration traffic; these migrations are ineffectual since the migrated blocks soon become cold as new blocks are written into the system. Colloid+ and Colloid++ perform more migration, resulting in worse performance. Cerberus efficiently balances the workload by dynamically allocating a portion of new writes to the capacity device when the performance device is saturated; as a result, Cerberus effectively uses the bandwidth of each device as the intensity increases.

\noindent \textbf{Summary.} For a variety of static workloads, Cerberus achieves better throughput than striping, Orthus, BATMAN, HeMem, and Colloid. We omit BATMAN in subsequent experiments since it performs worse than Colloid or its variants.

\subsection{Dynamic Workloads}
\label{sec:micro_dynamic}

\begin{figure}[t]
  \centering
  \begin{minipage}[b]{.49\linewidth}
    \centering
    \includegraphics[width=\linewidth]{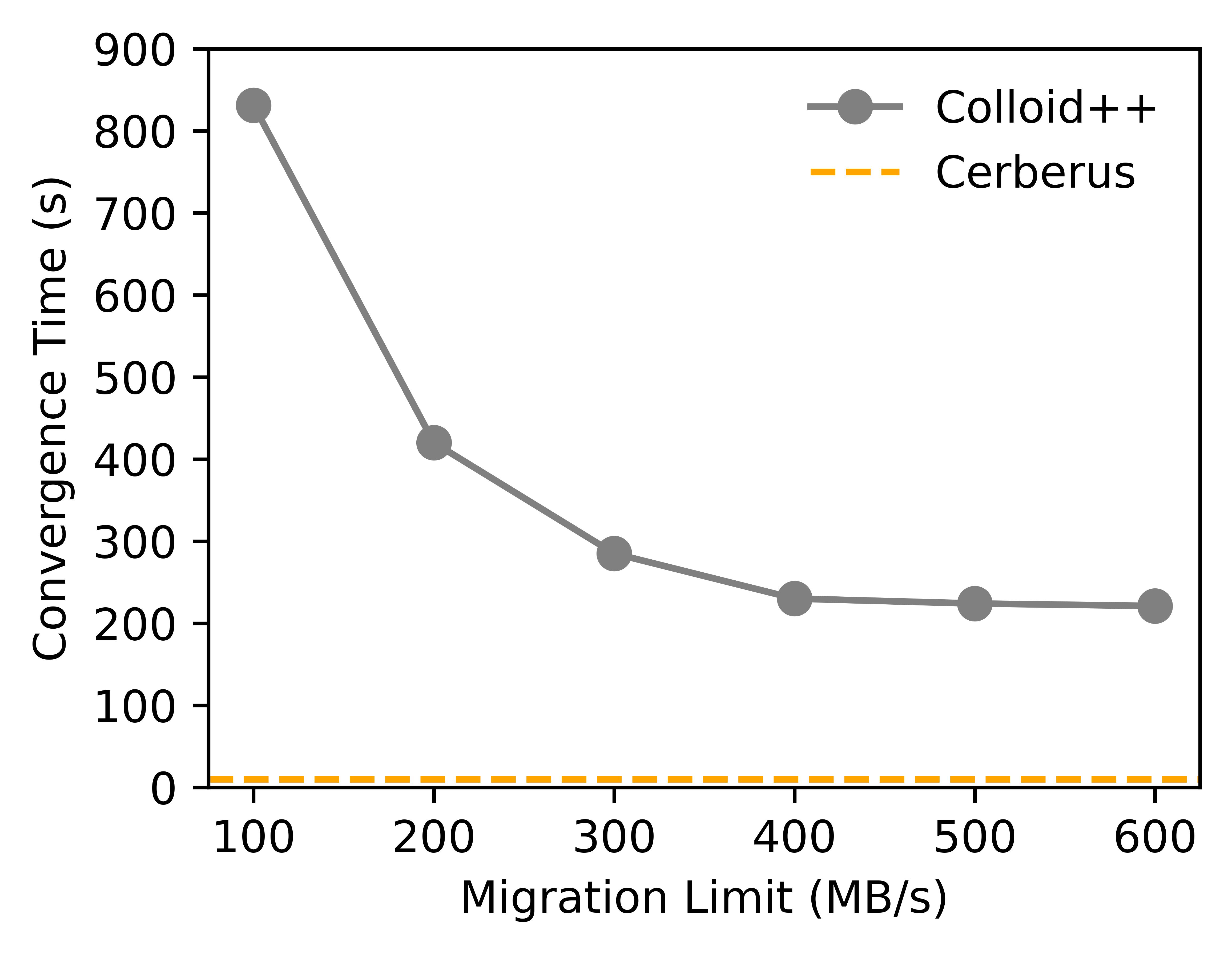}
     \subcaption{\textbf{Migration Limit}}
    \label{fig:dynamic_readonly_migration_limit}
  \end{minipage}%
  \begin{minipage}[b]{.49\linewidth}
    \centering
    \includegraphics[width=\linewidth]{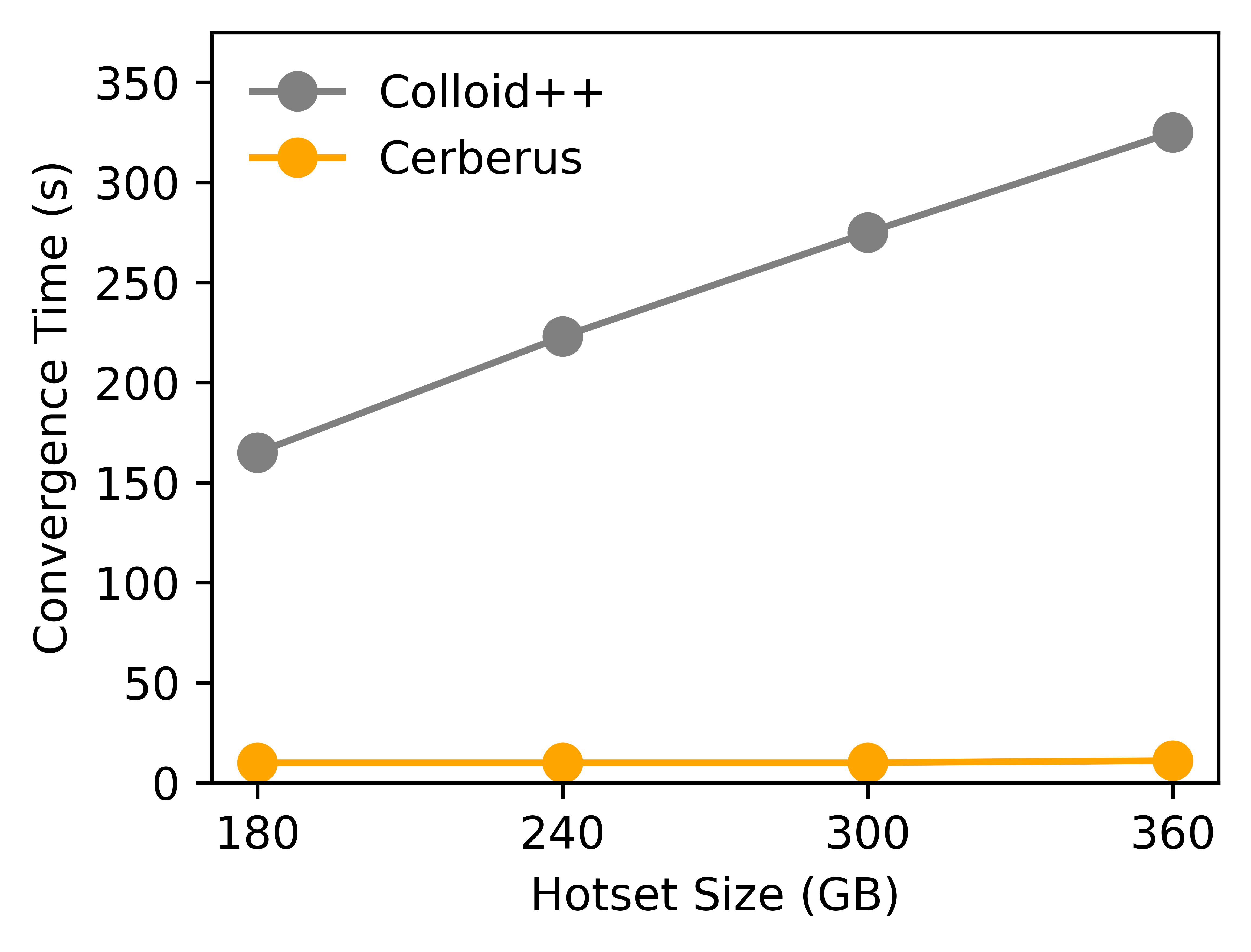}
    % \caption{\textbf{Read/Write Mixed Workload under Various Hierarchies}}
    \subcaption{\textbf{Hotset Size}}
    \label{fig:dynamic_readonly_hotset_conv}
  \end{minipage}
  \caption{\textbf{Limitation of Migration-based Approach.} \textit{The same workload as Figure \ref{fig:dynamic_readonly}.}}
  \vspace{-0.1in}
\end{figure}

Dynamic time-varying workloads are challenging for previous approaches.  To compare Cerberus to HeMem and Colloid++, we create a micro-benchmark modeling a bursty workload similar to those in practice \cite{berg2020cachelib}. After pre-warming the system with intensive load for 1000 seconds, the benchmark transitions to generating a 2-minute burst every 15 minutes. Figure~\ref{fig:micro_dynamic} shows the delivered throughput and amount of data promoted and demoted as a function of load (i.e., thread count) for read-only, write-only, and read-write workloads.

\noindent \textbf{Read-only workload.} Before 1000 seconds, the systems operate under high load; in this warm-up period, Colloid demotes and Cerberus mirrors approximately 7\% of the hot data to the capacity device. At 1000 seconds, when workload intensity drops, Colloid promotes the hotset back to the performance device, causing intensive promotion traffic (as shown in the second graph), which negatively impacts throughput (first graph). In contrast, within 10 seconds Cerberus efficiently rebalances traffic by routing requests to mirrored data back to the performance device.  When the 2-minute load burst occurs after about 1000 seconds, the performance device is saturated, causing its latency to surpass that of the capacity device. Colloid again demotes the hotset back to the capacity device (as depicted in the third graph). Consequently, Colloid performs worse than HeMem, which simply keeps hot data on the performance device and does not conduct any load balancing.  Again, during the load burst, Cerberus efficiently offloads excess traffic by redirecting requests for mirrored data back to the capacity device. Cerberus matches HeMem's performance under low load and achieves 1.53$\times$ higher throughput during workload bursts since Cerberus also utilizes the capacity device.

\noindent \textbf{Write-only and Read-Write Workloads.}
The other workloads in Figure \ref{fig:dynamic_writeonly} and Figure \ref{fig:dynamic_rwmixed_micro} demonstrate similar behavior.  Cerberus is able to load balance aligned 4KB writes by using subpages that track the valid versus invalid portions of each 2MB segment. For the write-only workload, Cerberus achieves 1.48$\times$ higher throughput than  HeMem under high load conditions.  
\begin{figure*}[t]
    \centering
    \begin{minipage}[b]{0.24\linewidth}
        \centering
        \includegraphics[width=\linewidth]{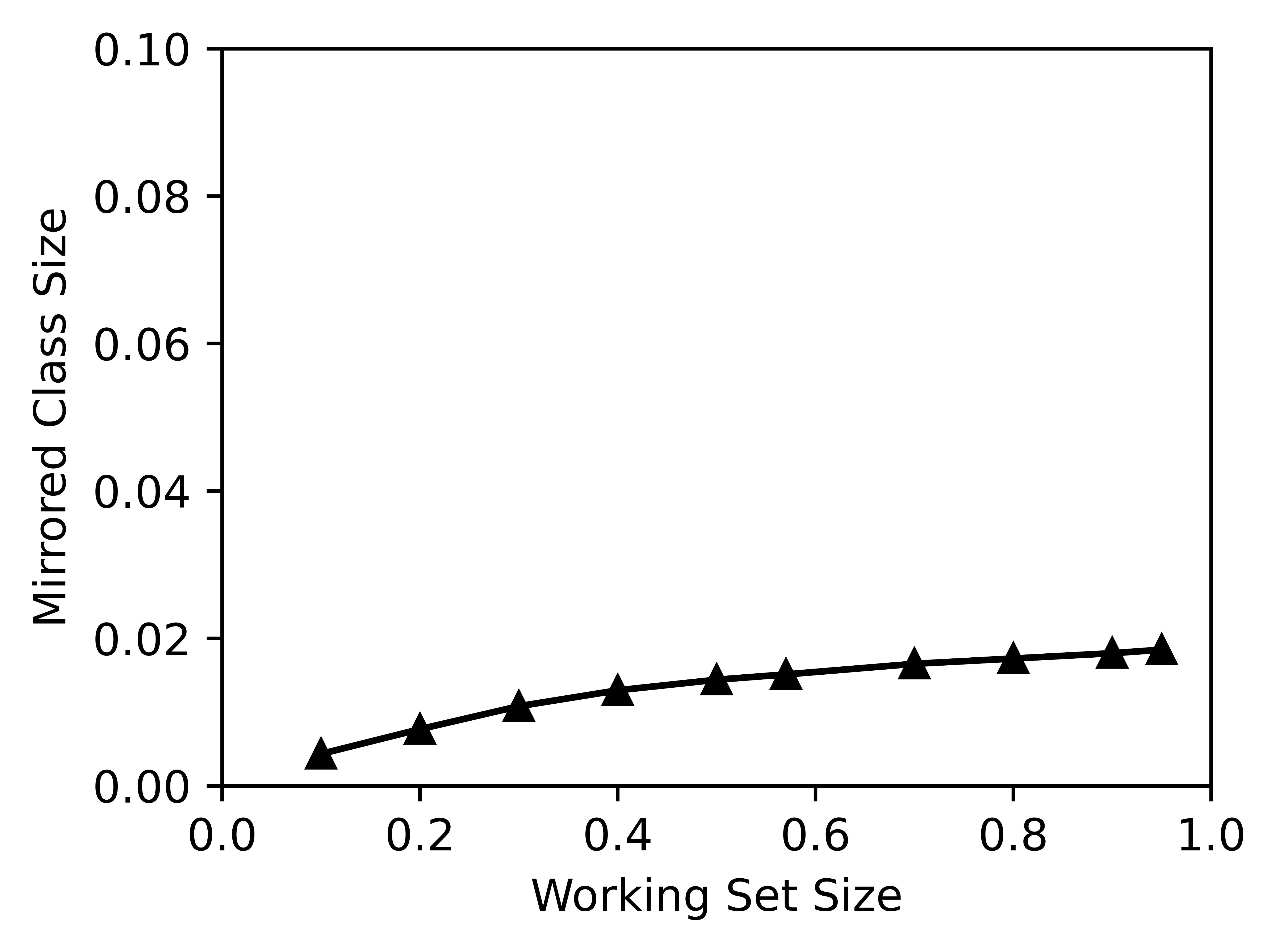}
        \subcaption{\textbf{Working set vs. Mirrored}}
        \label{fig:micro_static_ws_size}
        % \label{fig:static_writeonly}
    \end{minipage}
    \begin{minipage}[b]{0.24\linewidth}
        \centering
        \includegraphics[width=\linewidth]{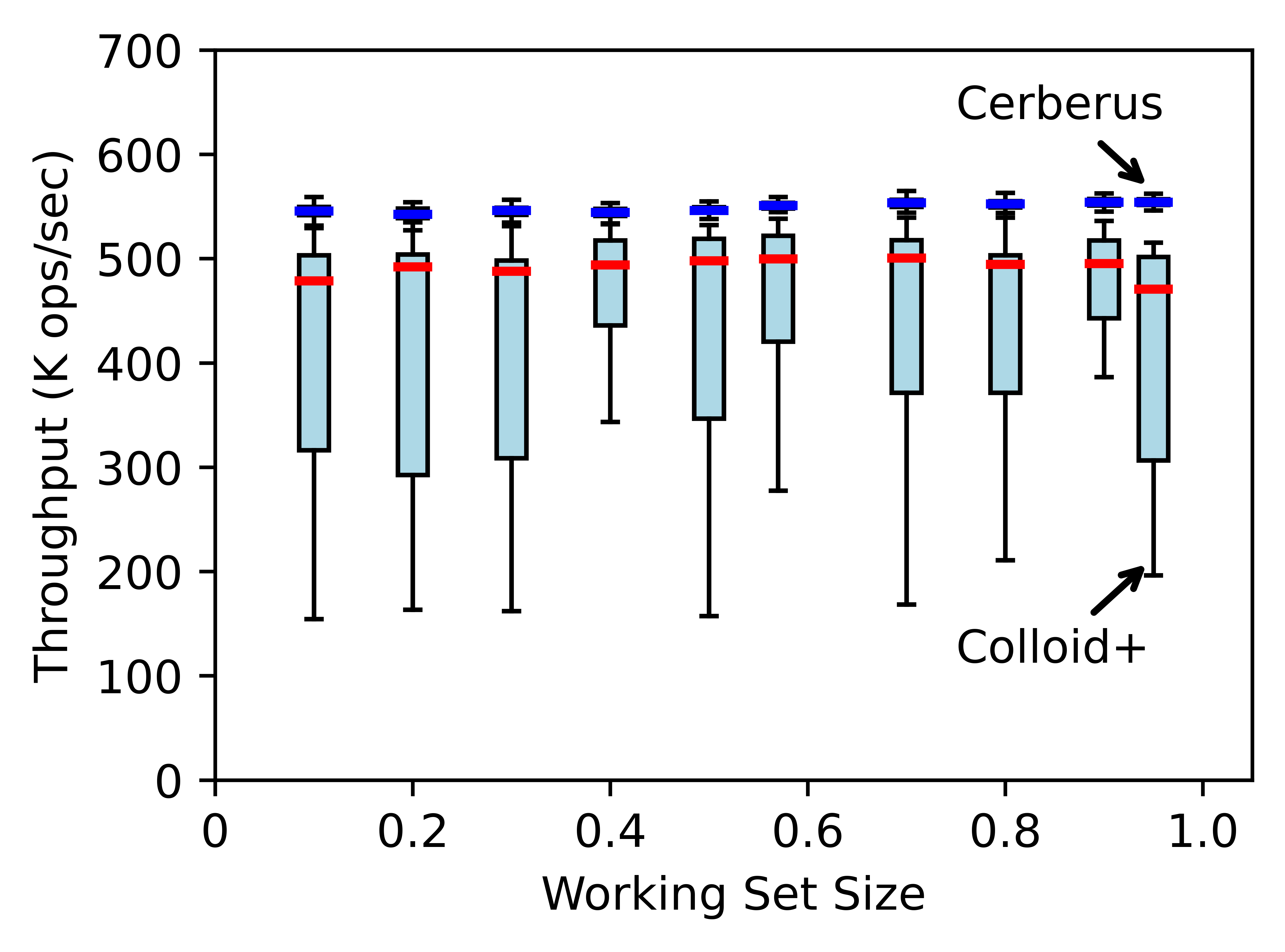}
        \subcaption{\textbf{Working set vs. Throughput.}}
        \label{fig:micro_static_ws_tp}
    \end{minipage}%
    \begin{minipage}[b]{0.24\linewidth}
        \centering
        \includegraphics[width=\linewidth]{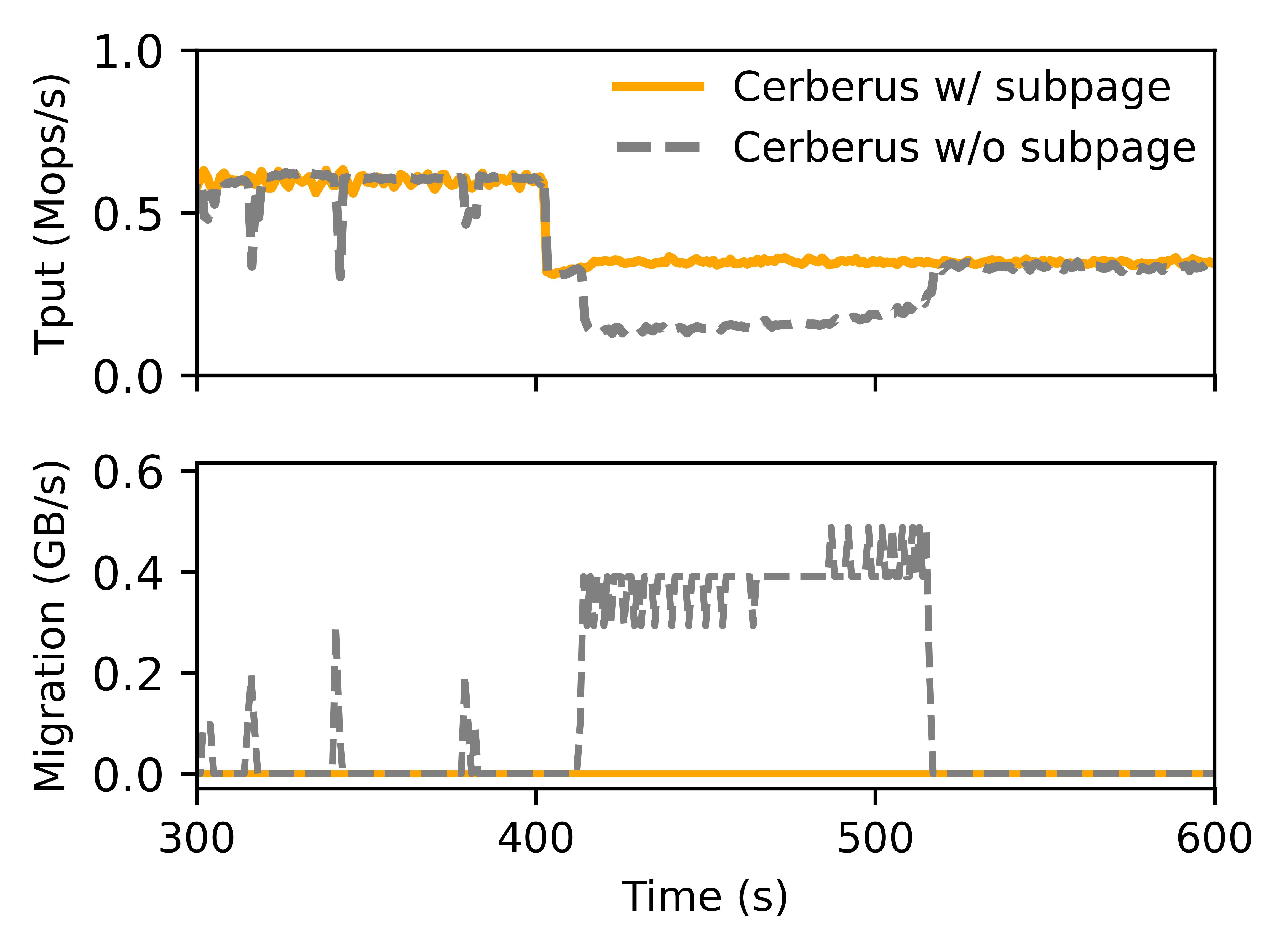}
        \subcaption{\textbf{Subpage Management.}}
        \label{fig:micro_static_subpage}
    \end{minipage}
    \begin{minipage}[b]{0.24\linewidth}
        \centering
        \includegraphics[width=\linewidth]{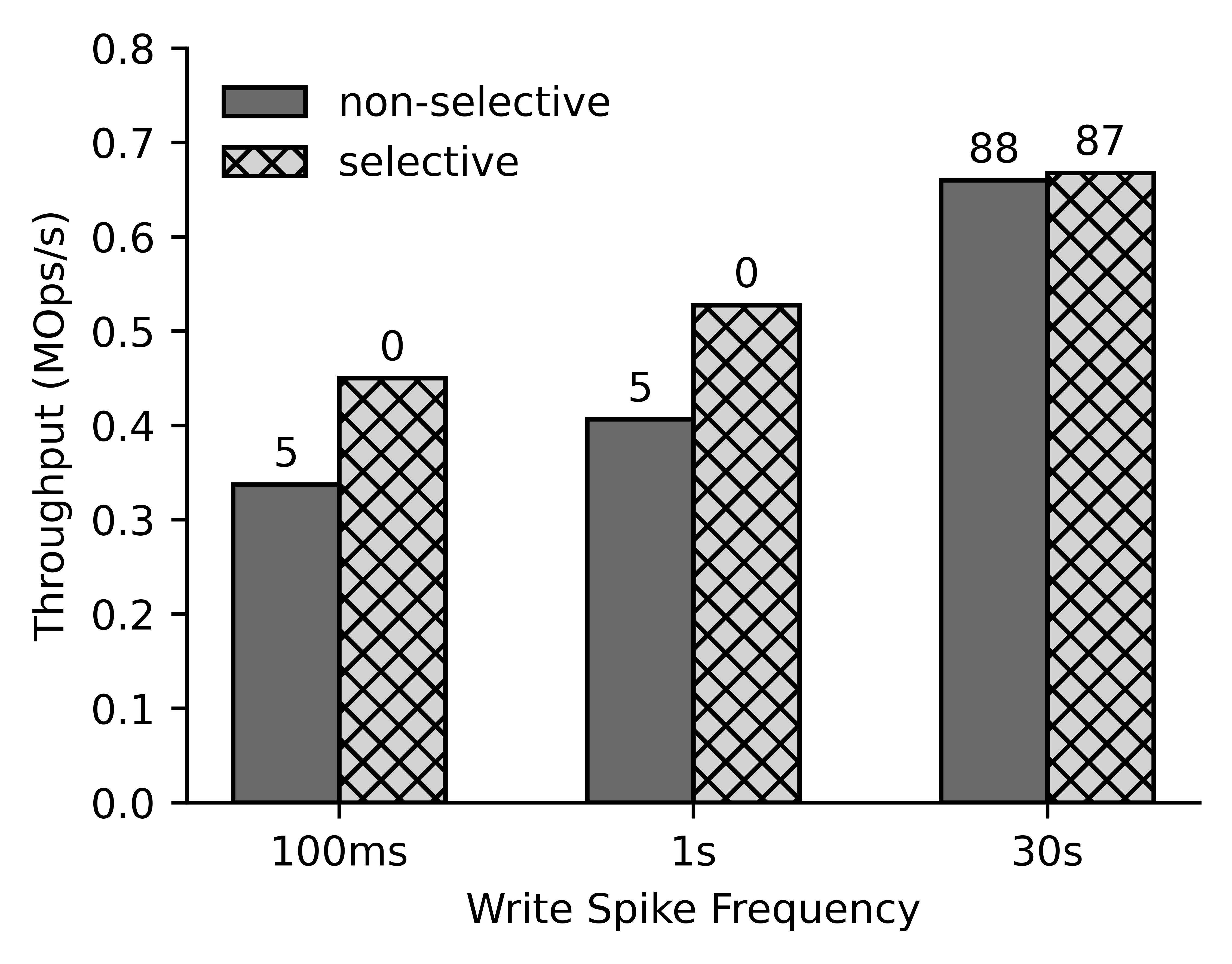}
        \subcaption{\textbf{Selective Cleaning.}}
        \label{fig:micro_static_cleaning}
    \end{minipage}
  \caption{\textbf{In-depth Analysis.}  \textit{In (a), workload is random read/write mixed  (50\% writes) at a high load (128 threads), varying the working set size, which is shown as a fraction of the total storage system capacity. The hotset is 20\% of the working set with 90\% access probability. In (d), the workloads are 256 threads workload with occasional write spikes every 0.1, 1, 30 seconds.}}
  \label{fig:static_ws}
  % \vspace{-0.6cm}
\end{figure*}

\begin{figure*}[t]
  \centering
  \begin{minipage}[b]{.5\textwidth}
    \centering
    \includegraphics[width=\linewidth]{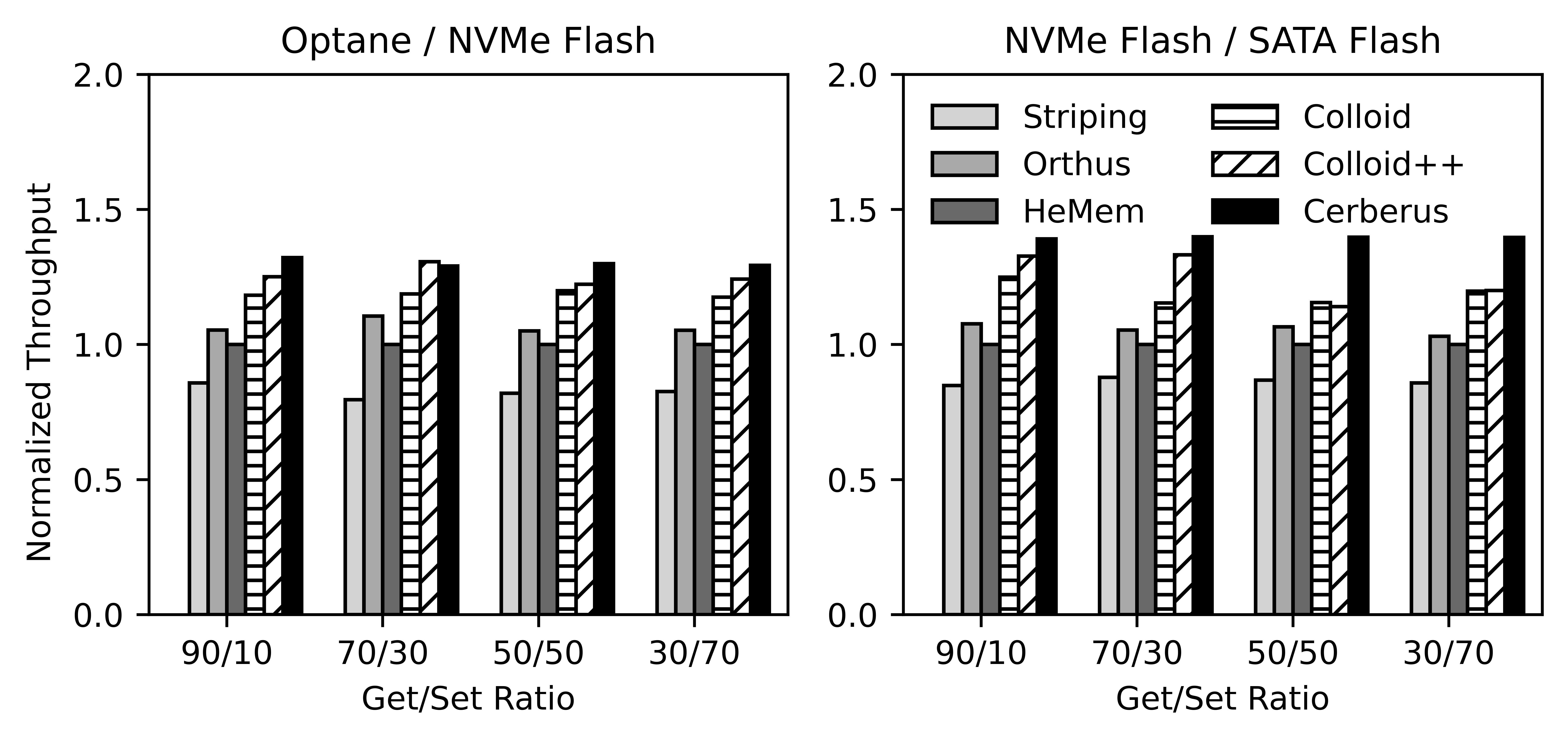}
     \subcaption{\textbf{Small Object Cache}}
    \label{fig:rwmixed_random}
  \end{minipage}%
  \begin{minipage}[b]{.5\textwidth}
    \centering
    \includegraphics[width=\linewidth]{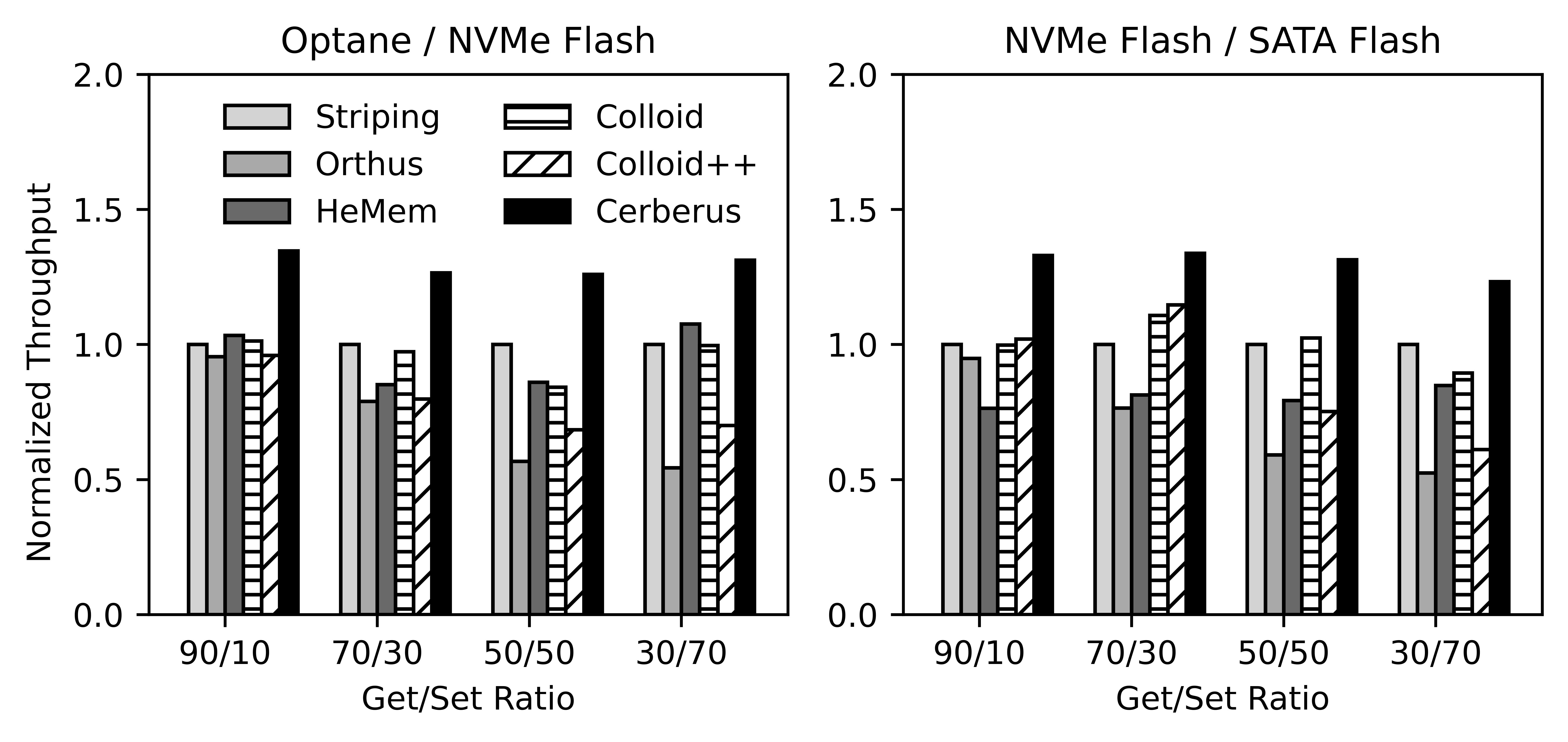}
    % \caption{\textbf{Read/Write Mixed Workload under Various Hierarchies}}
    \subcaption{\textbf{Large Object Cache}}
    \label{fig:rwmixed_log}
  \end{minipage}
  \caption{\textbf{Lookaside Cache Workload.} \textit{The DRAM cache is restricted to 200MB to stress each flash cache component in CacheLib. In (a), the workloads are 256 threads get/set-mixed Zipfian with 25M keys and 1KB value. SOC is set to 100GB, one third of the total capacity (100GB/200GB hierarchy). In (b), the workloads are 256 threads get/set-mixed Zipfian with 5M keys and 16KB value.}}
  \vspace{-0.1in}
\end{figure*}

\noindent \textbf{Disadvantages of Migrating Data.}
Colloid's reliance on migrating data has a number of disadvantages: not only does migration interfere with foreground traffic, but it also increases device writes and the time required for convergence. As shown in the caption of Figure~\ref{fig:micro_dynamic}, Colloid migrates on average 252\,GB to the performance and 229\,GB to capacity tiers across three representative workloads. In contrast, Cerberus only mirrors on average 86\,GB of data to the capacity tier. For instance, under a read-only workload, Colloid issues 282\,GB to the performance tier and 262\,GB to the capacity tier, while Cerberus writes just 87\,GB. Running this workload for one day results in 6.6\,DWPD and 3.1\,DWPD on the performance and capacity tiers, respectively. With a warranted endurance of 30\,DWPD over 5 years~\cite{intel2023optanessd}, the performance-tier device lasts 5.0 years with Cerberus. Colloid reduces the lifespan to 4.1 years, an 18\% drop. For the 1\,TB capacity-tier device rated for 0.37\,DWPD over 3 years~\cite{samsung2023nvme}, Colloid introduces 3.1\,DWPD of migration writes, shortening the lifespan from 3.0 years to 129 days, an 88\% reduction.

While Colloid's migration overhead can be controlled, limiting the rate of migration increases Colloid's convergence time, which is already significantly longer than Cerberus.  Figure \ref{fig:dynamic_readonly_migration_limit} shows Colloid's convergence time on the read-only workload when transitioning from low to high load given migration limits between 100MB/s (5 DWPD) to 600MB/s (30 DWPD). With a limit of 100MB/s, Colloid requires more than 800 seconds to adapt to a load increase, making it ineffective for workloads containing shorter burst intervals.  In contrast, Cerberus requires less than 10 seconds to adapt to load changes.  

Relying exclusively on migration to adapt to workload changes, also causes convergence time to increase as the size of the hotset increases.   
Figure \ref{fig:dynamic_readonly_hotset_conv} shows that Colloid's convergence time increases with larger (read-only) hotset since more data must be demoted to offload traffic from the performance device to the capacity device.  In contrast, Cerberus's convergence time is independent of the hotset size, since no migration is performed once the data is mirrored.

\noindent \textbf{Summary.} Cerberus integrates mirroring into classic tiering, allowing load adjustments without frequent migrations; this significantly enhances  adaptability to bursty workloads and reduces overall write traffic. In contrast, Colloid, and other tiering approaches, lack intra-tier redundancy and rely exclusively on migration for load balancing, since data can only be accessed from one location; under bursty workloads, Colloid incurs extensive writes for migration, greatly reducing device lifespan and resulting in performance worse than HeMem.

\subsection{Cerberus In-depth Analysis}
\label{sec:in_depth}
We now perform an in-depth analysis of Cerberus's techniques for determining the size of the mirrored class, tracking the state of subpages in mirrored data, and selective cleaning.

\noindent \textbf{Mirrored Class Size.} Cerberus performs efficient load balancing with a relatively small mirrored class. Figure~\ref{fig:micro_static_ws_size} shows the amount of space needed for Cerberus' mirrored class as a function of the workload's working set size.  Even when the working set size reaches 95\% of the total system capacity, Cerberus only mirrors 1.8\% of the total data.  Figure \ref{fig:micro_static_ws_tp} shows the corresponding throughput of Colloid+ and Cerberus.   The throughput of Colloid+ is highly unstable due to interference from frequent migrations; Cerberus consistently demonstrates higher and more stable throughput due to its effective use of a small amount of mirrored data.

\noindent \textbf{Mirrored Data Subpages.} The use of subpages within the mirrored class enables Cerberus to efficiently handle write requests.  Figure \ref{fig:micro_static_subpage} compares the behavior of Cerberus with subpages to Cerberus without subpages on a 4KB write-only workload undergoing a sudden load drop (from 128 to 8 threads) at 400 seconds.  After the load drop, Cerberus with subpages immediately redirects writes back to the performance device, quickly adapting to light load without requiring any data migrations. In contrast, Cerberus without subpages must migrate entire segments back to the performance device, as each write request is smaller than the segment size (2MB). Cerberus without subpages incurs additional migrations and significantly longer convergence times.

\noindent \textbf{Selective Cleaning.} Cleaning of mirrored data is rarely needed in Cerberus: cleaning is needed only for workloads containing a relatively small percentage of writes that invalidate a mirrored copy of data that is then frequently read (e.g., caching for ML models where write spikes occur when model parameters are refreshed \cite{yang2020analysis}).  Figure \ref{fig:micro_static_cleaning} shows the efficiency of Cerberus's selective cleaning under a read-intensive workload with occasional write spikes.  The number atop each bar represents the clean data percentage. We make two key observations. First, non-selective cleaning results in a 25\% decrease in cache throughput, but only a 5\% increase in the clean block percentage by cleaning those frequently written data. Second, Cerberus's selective cleaning policy effectively filters out data subject to high-frequency writing while cleaning long-term written data (writes every 30s).

\begin{table}[]

\resizebox{\columnwidth}{!}{%
\begin{tabular}{c|c|c|c|c|c|c}
Name           & Get  & Set  & LoneGet  & LoneSet & Key Size (B) & Avg Value Size (B) \\ \hline
A (flat-kvcache)    & 0.98 & 0    & 0.02     & 0       & 16-255       & 335              \\
B (graph-leader)   & 0.82 & 0    & 0.18     & 0       & 8-16         & 860             \\
C (kvcache-reg) & 0.87 & 0.12 & 1.04e-05 & 0.003   & 8-16         & 33112          \\
D (kvcache-wc)  & 0.6  & 0    & 8.2e-06  & 0.21    & 8-16         & 92422           
\end{tabular}
}

\caption{\textbf{Production Trace Distributions from CacheBench.} \textit{\textit{Flat-kvcache} and \textit{graph-leader} are from an application cache with small value sizes, resulting in mostly random traffic; \textit{kvcache-reg} and \textit{kvcache-wc} are from a storage cache with large value sizes, leading to log-structured traffic. LoneGet and LoneSet are requests for a key not present in the cache.}}

\label{table:real}
\vspace{-0.1in}
\end{table}

\subsection{CacheLib}
\label{sec:cachelib}
We now compare the end-to-end performance of CacheLib using Cerberus for its storage management layer, as compared to striping (CacheLib's default), caching (Orthus), and tiering (HeMem, Colloid and Colloid++).  For workloads, we use CacheBench\cite{meta2023cachebench}, a highly configurable benchmarking tool bundled with CacheLib that can model real-world production cache workloads~\cite{meta2023cachebenchreal}.  We also extend CacheBench to support Zipfian distributions and dynamic workloads.

\subsubsection{Static Workloads}
\label{sec:real-static}

We examine workloads that stress CacheLib's Small Object Cache (SOC) as well as its Large Object Cache (LOC).  

\noindent \textbf{Small Object Cache.}
We begin with 1KB lookaside cache workloads varying the Get/Set ratio across the two storage hierarchies. Figure \ref{fig:rwmixed_random} shows that striping, Orthus, and HeMem continue to deliver suboptimal performance.  Colloid and Colloid++ perform worse than Cerberus due to migration overhead caused by latency spikes, especially on the NVMe/SATA Flash capacity layer where NVMe and SATA exhibit more severe read/write interference than the Optane device.

\noindent \textbf{Large Object Cache.}  
The 16KB lookaside cache workload stresses the Large Object Cache, which stores large key-value pairs in a sequential log and an in-memory index; this results in sequential writes to the storage management layer and reads primarily to the most-recently-written blocks (i.e., the log head).  Figure~\ref{fig:rwmixed_log} shows that HeMem and Colloid cannot utilize the capacity device bandwidth once the performance device becomes saturated. Cerberus performs optimally across all workloads, achieving up to 1.36$\times$ higher throughput on Optane/NVMe and up to 1.54$\times$ higher throughput on NVMe/SATA. 

\noindent \textbf{CPU Overhead.} Under the 4KB random workload that stresses the mirroring mechanism most (Figure~\ref{fig:rwmixed_random}), Cerberus slightly increases CPU utilization (0--1.5\%) at 256 threads compared to the best-performing baseline, Colloid++, due to mirrored-page tracking and routing logic.

\label{sec:production}
\begin{figure}[t]
  \centering
  \includegraphics[width=\linewidth]{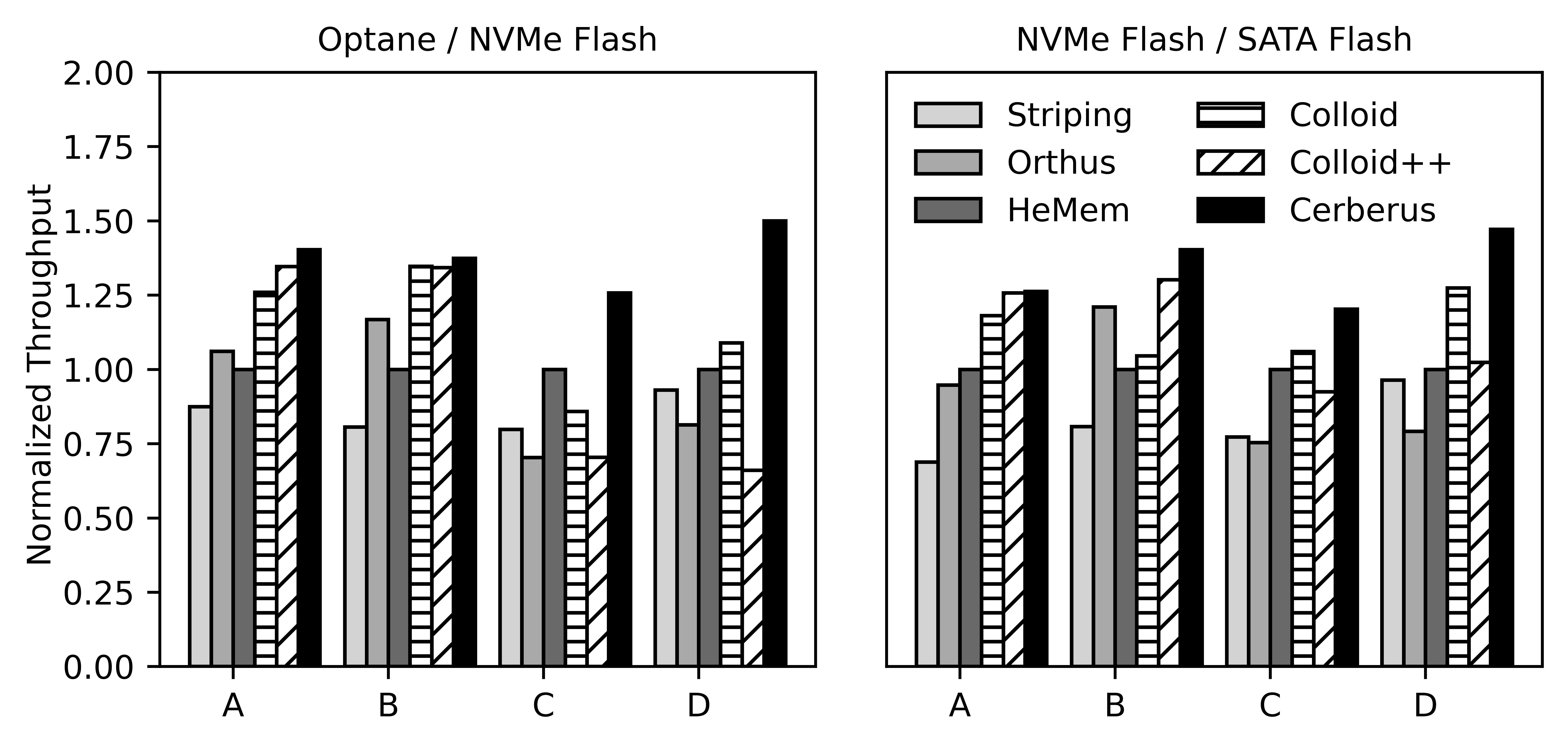}
  \caption{\textbf{Production Workloads.} \textit{DRAM cache is set to 1GB. For flat-kvcache and graph-leader, the SOC is set to one third of the total hierarchy capacity. We use 256 threads for flat-kvcache, graph-leader, and kvcache-wc and 80 threads for kvcache-reg. The cache throughput is normalized to HeMem's performance.}}
  \label{fig:real}
  \vspace{-0.1in}
\end{figure}

\begin{table}[t]
  \centering
  \footnotesize
  \resizebox{\columnwidth}{!}{
  \setlength{\tabcolsep}{1pt}
   
  \begin{tabular}{clccccccc}
  \toprule
  \textbf{Device} & \textbf{Workload} & \textbf{Metric} & \textbf{Striping} & \textbf{Orthus} & \textbf{HeMem} & \textbf{Colloid} & \textbf{Colloid++} & \textbf{Cerberus} \\
  \midrule
  \multirow{8}{*}{\parbox{1.2cm}{\centering\textbf{Optane}\\\textbf{NVMe}}} & \multirow{2}{*}{A} & Avg (ms)& 0.81 & 0.67 & 0.71 & 0.56 & 0.52 & \textbf{0.50} \\
   & & P99 (ms)& 15.76 & 18.18 & 19.24 & 9.59 & 9.00 & \textbf{7.33} \\
  \cmidrule{2-9}
   & \multirow{2}{*}{B} & Avg (ms)& 0.73 & 0.61 & 0.59 & 0.45 & 0.44 & \textbf{0.43} \\
   & & P99 (ms) & 7.00 & 14.28 & 12.83 & 4.12 & 4.12 & \textbf{3.65} \\
  \cmidrule{2-9}
   & \multirow{2}{*}{C} & Avg (ms)& 0.76 & 0.86 & 0.59 & 0.67 & 0.88 & \textbf{0.47} \\
   & & P99 (ms)& 9.53 & 10.69 & 6.50 & 7.78 & 8.85 & \textbf{5.43} \\
  \cmidrule{2-9}
   & \multirow{2}{*}{D} & Avg (ms)& 9.67 & 12.38 & 8.26 & 7.15 & 8.32 & \textbf{3.16} \\
   & & P99 (ms)& 67.68 & 272.00 & 73.63 & 72.93 & 86.32 & \textbf{27.76} \\
  \midrule
  \multirow{8}{*}{\parbox{1.2cm}{\centering\textbf{NVMe}\\\textbf{SATA}}} & \multirow{2}{*}{A} & Avg (ms) & 2.30 & 1.76 & 1.58 & 1.35 & 1.21 & \textbf{1.15} \\
   & & P99 (ms)& 46.43 & 27.96 & 22.98 & 25.98 & 21.31 & \textbf{20.40} \\
  \cmidrule{2-9}
   & \multirow{2}{*}{B} & Avg (ms) & 1.90 & 1.41 & 1.54 & 1.48 & 1.17 & \textbf{1.10} \\
   & & P99 (ms)& 23.40 & 16.84 & 11.59 & 14.15 & 9.34 & \textbf{8.54} \\
  \cmidrule{2-9}
   & \multirow{2}{*}{C} & Avg (ms)& 1.95 & 2.00 & 1.49 & 1.40 & 1.62 & \textbf{1.23} \\
   & & P99 (ms)& 18.18 & 18.42 & 12.35 & 16.50 & 18.50 & \textbf{12.31} \\
  \cmidrule{2-9}
   & \multirow{2}{*}{D} & Avg (ms)& 27.55 & 32.15 & 26.69 & 19.98 & 16.68 & \textbf{16.09} \\
   & & P99 (ms)& 239.86 & 364.20 & 160.10 & 158.07 & 170.29 & \textbf{101.85} \\
  \bottomrule
  \end{tabular}
  }
  \caption{\textbf{Average and P99 GET Latency (ms) of Production Workloads.}}
  \label{tab:latency_comparison}
  \vspace{-0.15in}
\end{table}

\subsubsection{Production Cache Workloads} 
We evaluate four long-running production cache workloads provided by Meta~\cite{meta2023cachebenchreal}, as detailed in Table \ref{table:real}. Figure \ref{fig:real} summarizes the results. For Workload A (lookaside cache workload) and Workload B (graph leader workload), Cerberus performs slightly better than Colloid and Colloid++, as these workloads involve small key-value pairs and stress the Small Object Cache.  On Workloads C and D, Cerberus significantly outperforms others, as these stress the Large Object Cache (LOC); Cerberus effectively distributes writes through its dynamic write allocation. Cerberus demonstrates better performance across all four production workloads due to its efficient mirroring-based load balancing and dynamic allocation. Compared to Colloid, Cerberus achieves an average throughput improvement of 1.24$\times$ on the Optane/NVMe hierarchy and 1.17$\times$ on the NVMe/SATA hierarchy.

\noindent \textbf{Latency.}  
Table~\ref{tab:latency_comparison} shows that Cerberus reduces average latency by 14\% and P99 latency by 19\% on average compared to the best-performing baseline. Striping performs worst on workloads A and B due to bottlenecks from the slower device, while Orthus performs worst on log-heavy workloads C and D, consistent with Figures~\ref{fig:static_sequential} and~\ref{fig:static_readrecent}. Performance gains are more substantial on the Optane/NVMe hierarchy, with 20\% lower average latency and 26\% lower P99 latency compared to the best-performing baseline. In comparison, the NVMe/SATA hierarchy shows smaller improvements, with 6.6\% and 12\% reductions respectively.

\subsubsection{Dynamic Workload}
\begin{figure}[t!]
  \centering
  \begin{minipage}[b]{\linewidth}
    \centering
    \includegraphics[width=\linewidth]{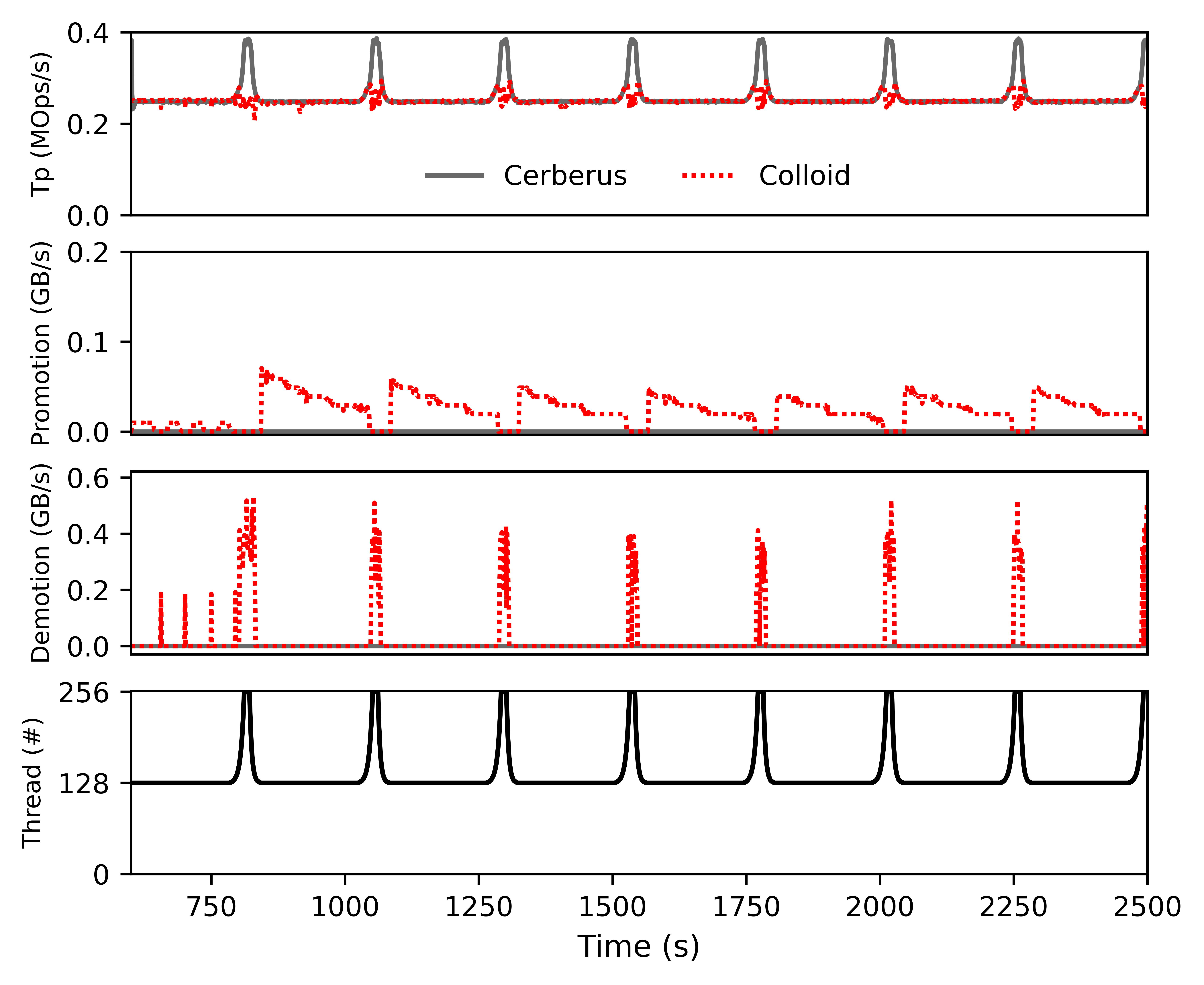}
  \end{minipage}
  \caption{\textbf{Dynamic Cache Workload.}  \textit{Read-heavy workload with 95\% GET and 5\% SET operations on Optane/NVMe hierarchy, where 20\% of keys belong to the hotset; the hotset is accessed uniformly at random with a 90\% probability. 25 million key-value pairs, with value sizes ranging from 2KB to 4KB; 1 billion operations. Size of SOC is 450GB.}}
  \label{fig:dynamic_real_readheavy}
  % \vspace{-0.05in}
\end{figure}
We compare how well Colloid and Cerberus handle load changes using the CacheBench benchmark \cite{meta2023cachebenchreal}. The bursts happen every 180 seconds and last for 60 seconds, matching those studied in \cite{berg2020cachelib} for data center workloads.  Figure \ref{fig:dynamic_real_readheavy} shows that Colloid struggles to adapt to these bursty workloads, generating significant migration traffic.  In contrast, Cerberus efficiently adapts to bursty workloads without incurring any migration overhead.  

\subsubsection{YCSB}
\begin{figure}[t!]
  \centering
  \begin{minipage}[b]{\linewidth}
    \centering
    \includegraphics[width=\linewidth]{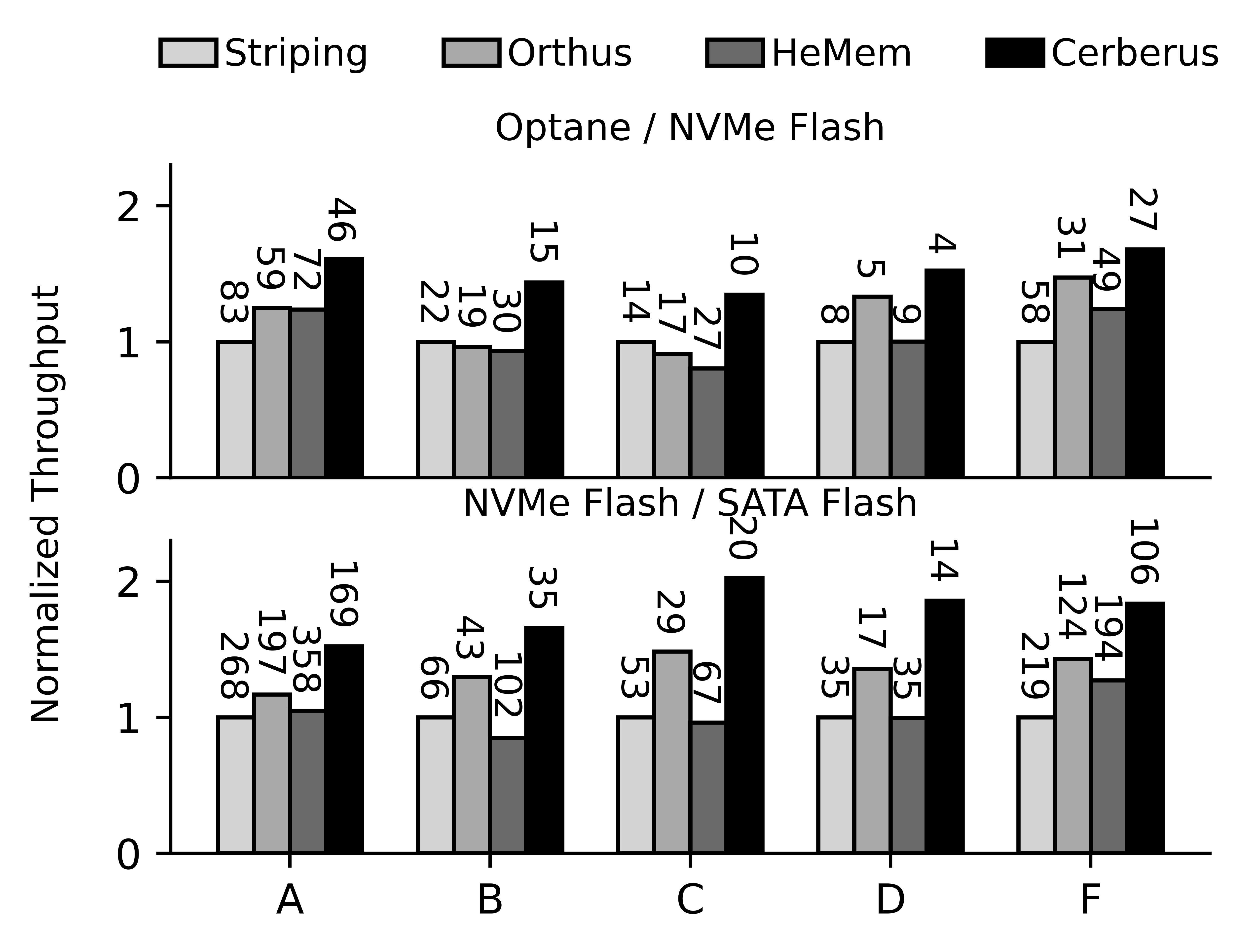}
  \end{minipage}
  \caption{\textbf{YCSB.} \textit{CacheLib uses the default parameters from CacheBench with 4GB DRAM cache. All the workloads are Zipfian ($\theta$ = 0.8) with 1KB values and 16-byte keys running under 256 threads with 20M records. The cache throughput is normalized to the throughput of the default (striping) system and the number above each bar is P99 latency in microseconds. Workload E is excluded due to CacheLib's lack of support for range queries.}}
  \label{fig:ycsb_static}
  \vspace{-0.1in}
\end{figure}

We compare Cerberus to striping, tiering (HeMem), and caching (Orthus) under the YCSB benchmark across two local hierarchies shown in Figure~\ref{fig:ycsb_static}. Since YCSB does not natively handle cache misses, we extended it to implement a lookaside caching pattern~\cite{berg2020cachelib}, where cache misses trigger a fetch from the backing store (simulated with a 1.5ms delay) and re-insertion into the cache. Cerberus shows up to $1.43\times$ higher throughput and 30\% less P99 latency compared to the best-performing system.

\section{Discussion}
\noindent \textbf{Multi-tier Extensions.} \Conceptname's mechanism naturally extends beyond the two-tier scenario. For example, data can be mirrored across multiple tiers, allowing requests to be dynamically routed to the tier with the lowest latency. This generalization calls for a more sophisticated optimization policy, which we leave as future work.

\noindent \textbf{Consistency.} \Conceptname can be extended to provide stronger consistency guarantees. One possible approach is to maintain a write-ahead log for mapping updates, such as those triggered by data migration. We leave a full study as future work.

\noindent \textbf{Performance Isolation.} Currently, \Conceptname transparently manages the underlying storage devices at the block level and is unaware of which tenant a given request belongs to. One potential solution is to use hints to associate each request with its corresponding tenant. With this additional metadata, \Conceptname can be extended to support and enforce performance isolation policies, such as fairness and quality of service (QoS), across multiple tenants.

\section{Related Work}
\noindent \textbf{Tiered Memory/Storage Systems.}
Our paper shares scope with research on tiered memory and storage~\cite{maruf2023tpp, raybuck2021hemem, dulloor2016heterogeneous, song2023prism, Raina2023prismDB, hoseinzadeh2019survey, abulila2019flat, Wang2014allocation, krish2014storage, zhang2010storage, wu2019autoscaling, lee2023memtis} for managing heterogeneous hierarchies of different devices, including DRAM, NVM, CXL-enabled memory, NVMe SSD, SATA SSD, and hard drives. vTMM~\cite{sha2023vtmm}, a tiered memory management system for virtual machines, optimizes memory-access tracking through page-modification logging. Nimble~\cite{yan2019nimble} is an OS-level system designed to improve page migration throughput. HeterOS~\cite{kannan2017heteros} functions as an OS/VMM-level heterogeneous memory manager, coordinating memory placement and migration with guest OSes. Thermostat~\cite{Agarwal2017thermostat} identifies hot memory pages using page table sampling. All of these systems are orthogonal to MOST, as they focus on memory-access tracking, hot/cold data identification, and data migration rather than load balancing across tiers.

Strata~\cite{Kwon2017strata} is a cross-media file system designed to mitigate the limitations of NVM, SSD, and HDD in a traditional storage hierarchy. Ziggurat~\cite{zheng2019ziggurat} migrates cold data to lower tiers in an NVM-disk tiered file system. Spitfire~\cite{zhou2021spitfire} is a three-tier caching-based buffer manager that leverages machine learning for optimal data placement. AutoRAID~\cite{wilkes1996hp} integrates mirroring for improved write performance with RAID-5 for cost-effective capacity. hStorage-DB~\cite{luo2012hstorage} enforces QoS policies for requests in hybrid storage systems. However, these systems are load-unaware and not able to fully utilize lower-tier bandwidth. PolyStore~\cite{ren2025polystore} is a meta layer built on top of storage medium-optimized file systems that maximizes cumulative storage bandwidth. However, PolyStore is optimized exclusively for file systems and does not maintain cross-tier duplication, therefore not able to efficiently adapt to dynamic workload.

Distributed multi-tier caching systems~\cite{rashmi2016ec, ren2024elect, liu2021emrc} broadly share a similar problem space with our work. However, their approaches differ significantly from MOST in both design and objectives. For instance, EC-CACHE~\cite{rashmi2016ec} uses erasure coding instead of mirroring and is limited to load balancing within a single tier. ELECT~\cite{ren2024elect} replicates data within a single tier and lacks support for cross-tier load balancing. eMRC~\cite{liu2021emrc} focuses on multi-tier miss ratio approximation but does not address the load balancing problem, making its objective orthogonal to ours. In contrast, MOST introduces a novel data management model that supports selective mirroring and adaptive load balancing across a heterogeneous storage hierarchy.

Machine learning-based storage and memory tiering approaches (e.g., IDT~\cite{oussama2024autonomous}, RLTiering~\cite{liu2022rltiering}, A3C~\cite{yi2025artmem}, ArtMem~\cite{chang2024idt}) target a similar problem space as our work. These systems typically utilize reinforcement learning to guide data placement decisions but are built on traditional single-copy tiering models. As shown in our evaluation of Colloid, such models struggle with dynamic and time-varying workloads. Like Colloid, these ML-based approaches are fundamentally constrained by the inherent limitations of single-copy designs and fail to address the critical challenge of load balancing across tiers in heterogeneous hierarchies.

\noindent \textbf{Storage Aware Caching/Tiering.} Our work aligns with research on storage-aware caching and tiering~\cite{wu2012novel, kim2018sib, ahmadian2019lbica, wu2021storage, qiu2023frozen, wu2022nyxcache}. Wu et al.~\cite{wu2012novel} identified a similar issue, where SSDs become throughput bottlenecks, and proposed migrating data to HDDs when SSD response times exceed those of HDDs. However, this migration-only approach reacts slowly to workload changes and incurs high migration overhead as Colloid. SIB~\cite{kim2018sib} targets HDFS clusters with SSDs and HDDs, using SSDs as write buffers and offloading reads to HDDs. LBICA~\cite{ahmadian2019lbica} implements cache load balancing by halting new allocations to performance devices under burst loads but does not balance read traffic. Like Orthus, these caching-based systems fail to fully utilize the capability of modern storage hierarchies.

% Their approach encounters limitations similar to BATMAN: they depend on migration to balance load, incurring significant migration costs for dynamic read-intensive workloads.

% \noindent \textbf{Flash Cache.} Our works builds upon research on flash cache \cite{flashshied, berg2020cachelib, mcallister2024fairywren, mcallister2021kangaroo,tang2015ripq}. These works are orthogonal to ours, as they focus on reducing write amplification in single-device flash setups, while we focus on multi-device management. We believe our work complements their designs by allowing them to scale seamlessly to multi-device hierarchies, as our system manages the underlying devices transparently.
\section{Conclusion}
We introduced \conceptnamefull, a new tier-based approach optimized for modern heterogeneous storage hierarchies. We demonstrate that with a small amount of mirrored data, \Conceptname improves throughput, reduces latency, enhances adaptability, and decreases migration-induced writes under dynamic workloads. In the future, hybrid approaches that integrate elements of caching, tiering, and RAID may offer further benefits, potentially leading toward a ``unified theory'' of storage hierarchy management.

\section{Acknowledgement}
We thank Huaicheng Li (our shepherd), the anonymous reviewers and the members of ADSL for their valuable input. This material was supported by funding from the NSF under the award number CNS-2402859 and the taxpayers of Wisconsin and the USA. We thank
NetApp, Microsoft, InfluxData, and GE HealthCare for their
generous support. Any opinions, findings, and conclusions or recommendations expressed in this material are those of the authors and may not reflect the views of NSF or any other institutions.
\newpage
%-------------------------------------------------------------------------------
\bibliographystyle{plain}
\bibliography{references}

%%%%%%%%%%%%%%%%%%%%%%%%%%%%%%%%%%%%%%%%%%%%%%%%%%%%%%%%%%%%%%%%%%%%%%%%%%%%%%%%
\end{document}